\documentclass[10pt]{article}

\usepackage{amsmath}
\usepackage{amssymb}

\usepackage{graphicx}

\usepackage{cite}

\usepackage{color} 

\usepackage[labelfont=bf,labelsep=period,justification=raggedright]{caption}

\makeatletter
\renewcommand{\@biblabel}[1]{\quad#1.}
\makeatother

\date{}

\definecolor{violet}{rgb}{1.00,0.00,1.00}
\definecolor{turquoise}{rgb}{0.00,0.40,0.50}    
\definecolor{lightred}{rgb}{0.9,0,0}		
\definecolor{brickred}{rgb}{0.8,0.5,0}
\definecolor{cadmiumgreen}{rgb}{0.0, 0.42, 0.24}

\begin{document}


\begin{centering}
{\it Frontiers in Fractal Physiology}, special issue
on ``{\it Critical Brain Dynamics}'' \\ (Edited by He BY,
Daffertshofer A, Boonstra TW), in press, 2012.
\ \\ \ \\ \ \\ \ \\ 
{\LARGE \textbf{Avalanche analysis from multi-electrode
ensemble recordings in cat, monkey and human cerebral cortex 
during wakefulness and sleep}}
\\ \ \\ \ \\ 
Nima Dehghani$^{1,\ast}$, 
Nicholas G. Hatsopoulos$^{2}$, 
Zach D. Haga$^{2}$,
Rebecca A. Parker$^{3}$,
Bradley Greger$^{4}$,
Eric Halgren$^{5}$, 
Sydney S. Cash $^{6}$, 
Alain Destexhe$^{1,\ast}$
\\ \ \\ 
\bf{1} Laboratory of Computational Neuroscience. Unit\'e de 
Neurosciences, Information et Complexit\'e (UNIC). CNRS. 
Gif-sur-Yvette, France.
\\
\bf{2} Committee on Computational Neuroscience and Department of
Organismal Biology and Anatomy, University of Chicago, Chicago, Illinois, USA.
\\
\bf{3} Interdepartmental Program in Neuroscience, University of Utah, Salt
Lake City, Utah, USA.
\\
\bf{4} Department of Bioengineering, University of Utah, Salt Lake City, Utah,
USA.
\\
\bf{5} Multimodal Imaging Laboratory, Departments of Neurosciences and
Radiology, University of California San Diego, La Jolla, CA, USA.
\\
\bf{6} Department of Neurology, Massachusetts General Hospital and Harvard Med.
School, Boston, MA, USA.
\\
$\ast$ E-mail: dehghani@unic.cnrs-gif.fr, destexhe@unic.cnrs-gif.fr \\
\end{centering}

\subsection*{Abstract}

Self-organized critical states are found in many natural systems,
from earthquakes to forest fires, they have also been
observed in neural systems, particularly, in neuronal
cultures. However, the presence of critical states in the awake
brain remains controversial. Here, we compared avalanche analyses
performed on different {\it in vivo} preparations during
wakefulness, slow-wave sleep and REM sleep, using
high-density electrode arrays in cat motor cortex (96 electrodes),
monkey motor cortex and premotor cortex and human
temporal cortex (96 electrodes) in epileptic patients. In neuronal avalanches
defined from units (up to $\>$160 single units), the size of
avalanches never clearly scaled as power-law, but rather scaled
exponentially or displayed intermediate scaling. We also analyzed
the dynamics of local field potentials (LFPs) and in particular LFP
negative peaks (nLFPs) among the different electrodes (up to 96
sites in temporal cortex or up to 128 sites in adjacent motor and
pre-motor cortices). In this case, the avalanches defined from
nLFPs displayed power-law scaling in double logarithmic
representations, as reported previously in monkey. However,
avalanche defined as positive LFP (pLFP) peaks, which are
less directly related to neuronal firing, also displayed
apparent power-law scaling. Closer examination of this scaling
using the more reliable cumulative distribution function (CDF) and
other rigorous statistical measures, did not confirm power-law
scaling. The same pattern was seen for cats, monkey and human, as
well as for different brain states of wakefulness and sleep. We
also tested other alternative distributions. Multiple
exponential fitting yielded optimal fits of the avalanche dynamics
with bi-exponential distributions. Collectively, these results
show no clear evidence for power-law scaling or self-organized
critical states in the awake and sleeping brain of mammals, from
cat to man.

\subsection*{keywords}
Criticality, Self-organization, Brain Dynamics, Scale invariance,
Complexity, Power-law


\section*{Introduction}

Self-organized criticality (SOC) is a dynamical state of a system
which maintains itself at (or close to) a phase transition point. 
This family of systems were initially described work by Bak, Tang
and Wiesenfeld (1987), and have been found in many natural systems
(reviewed in Bak, 1996; Jensen, 1998). SOC systems are
characterized by scale invariance, which is usually identified as a
power-law distribution of characteristics of the system's dynamics
such as event size or the waiting time between events.  The
temporal fingerprint of SOC systems is often $1/f$ or
$1/f^{2}$ noise. These features are interesting because
they show the presence of long-lasting or long-range correlations
in the system.

The dynamics of SOC systems are structured as ``avalanches''
of activity, separated by silent periods. Avalanche sizes are
typically distributed as a power law, where the probability of
occurrence $p(x)$ of a given avalanche size $x$ scales as: $$ p(x)
\sim x^{-\alpha} ~ , $$ where $\alpha$ is the scaling exponent of
the distribution.  

SOC systems have been observed in many different natural phenomena,
from sandpiles, to rice piles, in forest fires and earthquakes
(Frette, 1996; Malamud et al., 1998; Bak, 1996; Jensen 1998, Peters
2006). Evidence of SOC was also demonstrated in circuits of
neurons {\it in vitro} (Beggs and Plenz, 2003), where network
activity was found to alternate between active and quiescent
periods, forming ``neuronal avalanches''. The presence of
avalanches, although clear {\it in vitro}, is more controversial
{\it in vivo}. Since power-laws fit neuronal avalanches better than
other alternative probability distributions (Klaus and Plenz,
2011), their presence has been taken as evidence for neuronal
avalanches {\it in vivo}. In anesthetized cats (Hahn et al., 2010)
and awake monkeys (Peterman et al., 2009), power-law
distributed avalanches have been found in the peaks of local field
potentials (LFP). However, LFP peaks are only statistically
related to neuronal firing. In a study on awake and naturally
sleeping cats, no sign of avalanches were found in neuronal firing
(Bedard et al., 2006), and the apparent power-law scaling of LFP
peaks could be explained as an artifact induced by the thresholding
procedure used to detect LFP peaks. Previous studies have shown
that even purely stochastic processes can display power-law scaling
when subjected to similar thresholding procedures (Touboul and
Destexhe, 2010). It was also stressed that power-law statistics can
be generated by stochastic mechanisms other than SOC (Giesinger,
2001; Chialvo, 2010; Touboul and Destexhe, 2010). Similarly, if
exponentially growing processes are suddenly killed (or ``observed''), a
power-law at the tail ends will emerge (Reed $\&$ Hughes 2002). This case, would
be similar to a non-stationary Poisson processes, or combining Poisson processes
at different rates, a situation that is likely to happen in the
nervous system. Such scenarios can give rise to spurious power
laws.

These contrasting results correspond to different preparations and
recording techniques, single units or LFPs, or different species,
so that it is difficult to compare them. In the present paper, we
attempt to overcome these shortcomings by providing a systematic
analysis of both units and LFPs for different species and different
brain states.



\section*{Materials and Methods}

\subsection*{Recordings}

\subsubsection*{Cat} 

Recordings of local field potentials (LFPs) and action
potentials (APs) were obtained from motor cortex in 2 felines (M1
and approximately hindlimb region). Commercially obtained
96-electrode sputtered iridium oxide film arrays (Blackrock
Microsystems, Inc., Salt Lake City UT) were chronically implanted
and recordings were performed in the awake, unrestrained feline (as
described in Parker et al., 2011). Electrodes on the array were
arranged in a square with 400 micron spacing and 1 mm shank length.
LFPs and APs were recorded using a Cerebus data acquisition system
(Blackrock Microsystems). Spike sorting on AP data was performed
using the t-dist EM algorithm built into Offline Sorter (Plexon,
Inc.). All animal procedures were performed in accordance with
University of Utah Institutional Animal Care and Use Committee
guidelines.

We also compared these data with previously published
multi-electrode data on cat parietal cortex (Destexhe et al.,
1999). In this case, a linear array of 8 bipolar electrodes
(separated by 1~mm) was chronically implanted in cortical area 5-7,
together with myographic and oculographic recordings, to insure
that brain states were correctly discriminated (quiet wakefulness
with eyes-open, slow-wave sleep, REM sleep). Throughout the text, this cat will
be referred to as ``cat iii'' LFP signals were
digitized off-line at 250 Hz using the Igor software package
(Wavemetrics, Oregon; A/D board from GW Instruments, Massachusetts;
low pass filter of 100 Hz). Units were digitized off-line at 10
kHz, and spike sorting and discrimination was performed with the
DataWave software package (DataWave Technologies, Colorado; filters
were 300 Hz high-pass and 5 kHz low-pass).

\subsubsection*{Monkey} 

Recordings from three monkeys were used in this study. Each monkey was
chronically implanted with 100-electrode Utah arrays (400μm inter-electrode
separation, 1.0 mm electrode length; BlackRock Microsystems Inc., Salt Lake City
UT). In two monkeys (i) and (ii), we used recordings made during the
performance of motor tasks. The motor tasks involved moving a cursor to
visually-presented targets in the horizontal plane by flexing and extending the
shoulder and elbow of the arm contralateral to the cerebral hemisphere that was
implanted. In monkey (iii), sleep recordings were used
to test avalanche dynamics. Monkey i was implanted one 96 electrode array
in primary motor
cortex (MI) and a second 96 electrode array in dorsal premotor cortex
(PMd) from which
recordings were made on 64 electrodes in each cortical area. Monkey ii had an
array implanted in MI from which 96 electrodes were recorded. 
and Monkey iii had two arrays in MI and PMd from which 96 electrodes were
recorded in PMd cortex and 32 electrodes were recorded in MI area. During
a recording session, local field potential (LFP)
signals were amplified (gain, 5000), band-pass filtered (0.3 Hz to 250 Hz or 0.3
to 500 Hz), and recorded digitally (14-bit) at 1 kHz per channel To acquire
extracellular action potentials, signals were amplified (gain, 5000), band-pass
filtered (250 Hz to 7.5 kHz) and sampled at 30 kHz per channel. For each
channel, a threshold was set above the noise band: if the signal crossed the
threshold, a 1.6ms duration of the signal - as to yield 48 samples given a
sampling frequency of 30 kHz - was sampled around the occurrence of the
threshold crossing and spike-sorted using Offline Sorter (Plexon, Inc., Dallas,
TX). All of the surgical and behavioral procedures performed on the
non-human primates were approved by the University of Chicago’s IACUC and
conform to the principles outlined in the Guide for the Care and Use of
Laboratory Animals (NIH publication no. 86-23, revised 1985).

\subsubsection*{Human}  
Recordings were obtained from two patients with medically
intractable focal epilepsy using NeuroPort electrode array as
discussed previously (Peyrache et al., 2012, Truccolo et al.,
2010). The array, 1mm in length, was
placed in layers II/III of the middle  temporal gyrus with informed consent of
the patient and with  approval of the local Institutional Review Board in
accordance with  the ethical standards of the Declaration of Helsinki. This
array is silicon-based, made up of 96 microelectrodes with
400-$\mu$m spacing, covering an area of 4 $\times$ 4 mm. Since the
corners are omitted from the array, the furthest separated contacts
are 4.6 mm apart. Data were sampled at 30 kHz (Blackrock
Microsystems, Salt Lake City, Utah, USA). The continuous recording
was downsampled to 1250 Hz to obtain LFPs. The dataset we
analyzed was devoid of any form of identifiable epileptic activity
(such as interictal spikes), and there was no seizure in the
analyzed dataset. The implantation site was included in the
therapeutic resection in both patients. For details on spike
sorting, see Peyrache et al.\ (2012).  

\subsection*{Avalanche detection}

Avalanches are defined by temporally contiguous clusters of
activity among the different electrodes, separated by periods of
silence. Either trains of neuronal action potentials (spikes) or
LFP peaks can can be analyzed for the occurrence of avalanches.
There are two empirical limits
on bin duration. The smallest bin size is set by the duration of the action
potential. The upper boundary, is limited by how many
unique values of the aggregate ensemble activity occur in a window. When the
number of unique values approaches 1, avalanche
loses its definition, because there is no silent period left. In the cat data,
where there are $\>$160 cells, we reach
this limit at a bin-width of 16 ms. So, we have stayed within
the 1-15 ms regime in which an avalanche could be well defined.

\subsubsection*{Spike avalanche}

In each set of recordings, regardless of the spatial location of a
given electrode in the multielectrode array, its spiking activity
was put in the same pool with all other spikes recorded from other
electrodes of the same array. This ensemble trace was then binned
and coarse grained for different $\delta$t ranging from 1 ms to 16
ms in 2 ms steps. This created a series of bins containing the
ensemble of activity across all neurons for that
$\delta$t. The sum of spiking in that bin represents the total bin activity.
The sum of all bin activities between two quiescent bins,
represents the avalanche size, which was later used for statistical
analyses. Notice that in the case of the minimum $\delta$t = 1,
avalanche size would range between 0 and maximum number of neurons
present as this bin approximates the size unity of spiking period. 
Figure~\ref{fig1}A shows the definition of avalanche in spike
series from human recordings.

\subsubsection*{LFP avalanche} 

Each LFP trace was first detrended through a least-squares fit of a
straight line to the data and subsequent subtraction of the
resulting function from all the sample points. After this
detrending removed the mean value or linear trend from a LFP
vector, it was then normalized (Z score) to have a common reference
frame for discretization across channels, recordings, states and
species. The z-scored LFP, was then discretized through a local
maxima peak detection. An optimizing small running average filter
was designed and 3  passes of the filter were applied to the data in
order to remove small spurious peaks in each LFP deflection. 
Next, by comparing each element of data to its neighboring values,
if that sample of data was larger than both of its adjacent ones,
that element was considered as a local peak. Next, all the peaks
were sorted in descending order, beginning with the largest peak,
and all identified peaks not separated by more than minimum peak
distance (of 3 samples) from the next local peak were discarded. 

The threshold was fixed and defined as a multiple of the standard
deviation (STD) of the LFP signal.  Different thresholds were
tested, starting at 1.25$\times$STD and increasing in 0.25 steps up
to 5$\times$STD for both negative and positive maxima. This
procedure was realized on each LFP channel, state, species
(Fig.~\ref{fig1}B).  Such matrix of discrete events (for a given
polarity and a given threshold), was then treated the same way the
spike matrix was used to create avalanche vectors of quiescent and
active periods.

\subsection*{LFP peak and spiking relationship}

\subsubsection*{Wave-triggered-average (WTA)}

We used wave-triggered averaging (WTA) to analyze the differences
in the relationships of spikes to nLFP versus pLFP. In WTA, the
individual negative LFP peaks (nLFP) were used to epoch the
ensemble spike series. The epoched ensemble spike series were
normalized by the number of epochs (triggered by nLFPs). This
procedure was performed for the three different thresholds (low,
medium and high) and the results were averaged across these
thresholds to obtain cross-threshold WTA percentage firing to
quantify the spike-nLFP relationship.  An identical procedure was
applied to pLFPs. The red and blue solid lines in Figure~\ref{fig6}
refer to nLFP-spike and pLFP-spike WTA percentage firing,
respectively.

\subsubsection*{Controls and Randomization Methods}
We used 4 methods of surrogate/randomization in order to evaluate the
statistical robustness of the comparative relation of spike-nLFP vs spike-pLFP.
Each of the following 4 methods, was first performed on all 3 chosen thresholds
and then the results were averaged to obtain the overall randomization effect. 

\subparagraph*{{\it Poisson surrogate data}}
At the first step, we wanted to test whether the observed nLFP and pLFP
differences could be reproduced by surrogate spike series. For this type of
control, first, each individual channel's spike rate was calculated. Then, using
a renewal process, a surrogate Poisson spike series for that
channel was created (matching the firing rate and duration of the experimental
data from that channel). Then, all Poisson spike series (across all channels)
were aggregated together to create the ensemble spike series (similar to the
experimental data). Next, for each pLFP (or nLFP), the WTA of this Poisson
aggregate series was created. This procedure was repeated 1000 times and then
averaged across the 1000 trials. The results were close to a constant WTA
percent firing and did not fluctuate according to the timing of the peak LFP
that was used to epoch each individual WTA event. This control test showed that
the simple aggregate of surrogate Poisson spikes can not re-produce the observed
relation between nLFP and spikes in the WTA or mimic the behavior of natural
peak(positive or negative)-induced percentage firing. This procedure was also
repeated with Poisson spikes without a refractory period and provided similar
results.

\subparagraph*{{\it Random permutation}}

In a follow up test, we wanted to verify that randomizing the aggregate
spike
series by itself can not mimic the observed the LFP-spike relation. For
this procedure, we performed a random permutation on the aggregate spike series
and then calculated the nLFP(and pLFP)-based WTA. This procedure was repeated
1000 times. The observations are similar to the Poisson randomization, verifying
that the nLFP peak is not reproducible by randomization of spikes and the
fluctuations of WTA percentage firing are not results of random events.

\subparagraph*{{\it Local jitter randomization of LFP peaks}}
Next, we wanted to evaluate the effects of randomization based on the
statistics of the individual channel's LFP peak times (before aggregating them
into the ensemble LFP peak train). First, each channel's nLFP IPI
(inter-peak-interval) were calculated. Then these IPIs from all channels were
put in the same pool and the, 0.25, 0.5 and 0.75 quantiles IPI for the
aggregate nLFPs were extracted. Next, we created a normal distribution with 0.5
percentile as the mean, the interquartile range (0.75 quantile minus 0.25
quantile) as the standard deviation of the pdf, and N events matching the
number of aggregate nLFPs. This set of values, were used to jitter nLFPs in the
following manner. Each sample from the aggregates nLFP peak series was
shifted according to one drawn sample (without replacement) from the nLFP jitter
pool. The direction of the shift was to the right if the drawn jitter value was
negative (and to the left for the positive value). The magnitude of the shift
was defined by the value of the jitter itself. The same procedure was repeated
for pLFPs. The results of this randomization are shown in Figure~\ref{fig6}.A.
As can be appreciated, with this tightly regulated data-driven local
randomization, the structure of the WTA is preserved except for the peak curve
around 0 for the nLFP case.

\subparagraph*{{\it Fixed-ISI circular shift of spikes}}
In this procedure, we kept the ISI (inter-spike interval) of the aggregate
spike series as well as the IPI (inter-peak intervals) of the nLFP and pLFP
intact but randomized the relation between the aggregate spike and aggregate
peak series. In each of the 1000 trials, a circular shift with the magnitude
chosen randomly between 1 and the range of the ISI, was performed. The results,
shown in Figure~\ref{fig6}.B, show that by destroying the relation
between ensemble spikes and ensemble peaks while preserving their internal
structure, the observed fluctuations and most importantly, the tightly bound
relation of nLFP and spikes, is lost.

\subsection*{Testing power law distribution in empirical data}

For testing the power-law behavior, usually a simple least square
method is applied to fit a power-law on the data. If such fit in a
log-log scale, follows a straight line, the slope of the probability density
function (PDF) line
is taken as the scaling exponent. Such method is widely practiced
but is highly inaccurate in its estimation of true existence of
power-law in a given dataset. It has been argued that, for
obtaining statistically sound results in estimating power-law in
empirical data, one has to rely on rigorous statistical methods. 
In a detailed analysis of the problem (Newman, 2005; Clauset et
al., 2009), it was proposed that the cumulative distribution
function (CDF) is much more accurate to fit the power-law exponent,
as well as to identify if the system obeys a power-law. 

If the initial distribution of the PDF is
power-law, i.e., $$ p(x)  = Cx^{-\alpha} ~ , $$ then CDF is defined as $$
Pr(X>x) = C \int_{x}^{\inf} {x}'^{-\alpha} d{{x}'} =
\frac{C}{{\alpha}-1}x^{-({\alpha}-1)} ~ . $$ Thus, the
corresponding CDF also behaves as a power-law, but with a smaller
exponent $${\alpha}-1$$ being 1 unit smaller than the original
exponent (Newman, 2005).  

Generally, in fitting the power-law to the empirical data, all the
initial values (left hand of the distribution histogram i.e,
smallest sizes of avalanches) are included in the used decades to
obtain the slope of the fit (scaling exponent ${\alpha}$). The
inclusion of these initial parts may cause significant errors, and
should be removed (Clauset et al., 2009; Bauke, 2007; Goldstein et
al., 2004). Thus, before calculating the scaling exponent, it is
essential to discard the values below the lower bound ($X_{min}$). 
It is only above this lower bound that, a linear PDF or CDF
can be reliably used for estimation of the scaling exponent. 
There are different methods for proper estimation of the $X_{min}$.  We
used a Kolmogorov-Smirnov (KS test) optimization approach that
searches for the minimum ``distance'' (D) between the power-law
model and the empirical, where  for Xi$\geqslant$$X_{min}$, ``D'' is
defined as $$D = max | S(x) - P(x)| ~ ,$$ S(x) the CDF of the
empirical data and P(x) the CDF of the best matching power-law
model. The $X_{min}$ value that yields the minimum D, is the optimal
$X_{min}$. The $X_{min}$ is used in a maximum likelihood estimate (MLE) of
power-law fit to the CDF of the avalanches in order to obtain the
scaling exponent. This fitting, however, does not provide any
statistical significance on whether the power-law is a plausible fit
to the data or not. After the estimation of $X_{min}$ and the exponent,
we generated N (N=1000) power-law distributed surrogate data with
the exact same features of $X_{min}$ and exponent. Each of these
surrogate series are then fitted with power-law and KS statistics
of distance D (to the surrogate power-law), is performed. The
fraction of N that the resultant statistics was bigger than the one
obtained from the empirical data, comprises the p-value. If p-value
$\geqslant$ 0.1, the power-law is ruled out. However, even if
p-value is larger than this threshold, the data is not necessarily
guaranteed to be generated by a power-law process unless no better
distribution is found to estimate the properties of the data. For
this, the alternative test was adapted as following.

\subsubsection*{Generating power-law distributed random numbers with high
precision}

It is essential to use high precision and reliable algorithms to generate
random numbers from a given probability distribution; otherwise the statistical
tests based on such distributions may be erroneous. For initializing the
generator with an ``Integer Seed'', we adapted the reliable Mersenne Twister
algorithm (known as
MT19937AR) with full precision of Mersenne prime $(2^{19937}-1)$ (Mersenne et
al,. 1998). This algorithm provides a proper method for running Monte Carlo
simulations. After initialization, ``Transformation algorithm'' was
used to generate the desired distribution (Clauset et al, 2009, Press et al.,
1992a). All the random number generations and analyses were performed on a
16-core Intel 48 GB Linux platform equipped with 448 core Tesla C2050 GPU with
double precision of 515 Gflop and single precision of 1.03 Tflops. The custom
code was based on Matlab (Mathworks) and CUDA (NVIDIA) wrapper Jacket
(Accelreyes) for parallel computing on GPU.

\subsection*{Alternative fits}

The power-law fit was compared with alternative hypotheses to test
which distribution best fits the data.  The alternatives included
exponential distribution (as predicted by a Poisson type stochastic
process), "Discretized log-normal distribution" (which is
represented as a linear fit in log-normal scale), as well as fit of
"Discrete exponential distribution" nature. These fits had two
general types of simple exponential, defined as: $f(x) = a \exp{(b
x)}$ as well as sum of exponential set as: $f(x) = a \exp{(b x)} +
c \exp{(d x)}$ In each case, residual analyses, goodness of fit as
well as confidence and prediction bounds were used to evaluate the
properties of each fit vs power-law. In case of a good fit model,
Residual, defined as the difference between data and fit, should
approximate random error and behave randomly. 

\subsubsection*{Goodness of fit comparison of exponential models}

A measure of ``goodness of fit'', R-square, is the ratio of the sum
of squares of the regression (SSR) and the total sum of squares (SST).
This measure, represents the square of the correlation between the observed and
predicted response values, and indicates what percentage of the variance of the
data is explained by the chosen fit (values of R-square range from 0, worst fit,
to 1, the best possible fit). 
If we have SSR as: 
$SS_{\text{reg}} = \sum_i (\hat{y_i} -\bar{y})^2$,
and SSE as:
$SS_{\text{err}} = \sum_i (y_i -\hat{y})^2$,
and SST as:
$SS_{\text{tot}} = \sum_i (y_i-\bar{y})^2$,
where, $y_i ,\bar{y}, \hat{y}$ are the original data values, their mean and
modeled values respectively. Then, it follows that:
$$R^{2} = SS_{\rm reg} / SS_{\rm tot} =  1 - {SS_{\rm err}\over SS_{\rm tot}}.$$
\\Correction by ``total degree of freedom'' and ``error degree of freedom'',
defines adjusted R-square:
$$\bar R^2 = {1-(1-R^{2}){N-1 \over N-M-1}} = {1-{SS_\text{err} \over
SS_\text{tot}}}{df_t \over df_e}.$$
\\where ``N'' is the sample size, and ``M'' is the number of fitted
coefficients (excluding constants). Usage of $\bar R^2$ in the comparison of
``simple exponential'' and ``sum of exponential'' is warranted by the fact that
by an increase in the fitted number of the components, from one model to the
other, the degrees of freedom changes. Both $R^2$ and $\bar R^2$
measures were estimated through nonlinear least square optimization of
exponential curve fitting. In the optimization process for estimating the
coefficients of the models, we adapted Levenberg-Marquardt algorithm with a
tolerance of $10^-8$ (Press et al., 1992b). 

\subsubsection*{Test of linearity in log-normal scale} 

Linearity in log-normal scale, is a hallmark of an exponential family
process. 
In order to test the linearity of the PDF in log-normal scaling, we used Root
mean square error (RMSE),
$\operatorname{RMSE}(\hat{\theta}) = \sqrt{\operatorname{MSE}(\hat{\theta})}$
where MSE is:
$SS_{\rm err} \over df_e$. 
This measure ranges from 0 to 1, where closer value to 0 is an indicator of
a better fit. 
\\This test was performed by fitting $y = Log(P(x))$ with a linear
least square first degree polynomial. As shown in Fig.~\ref{fig13}C, sometimes,
the initial values in the left tail may slightly deviate from a simple 1st
degree polynomial. Therefore, we tested whether the linearity was
improved or worsened when the data range was reduced to above some $X_{min}$.
For doing so, we adapted a more stringent regression, bisquare robust 1st degree
polynomial (Press et al., 1992b). This method is an iteratively reweighted
least-squares, based on $\bar R^2$, and assigns less weight to the values
farther from the line. This procedure was repeated after
excluding consequent
single values from the left tail (up to 20 percent of the points). For each new
shortened series, the RMSE (based on bisquare method) was re-calculated. The
rational behind using RMSE for testing the linearity range in these datasets
(with variable N) is that when a distinct point is removed from the dataset, 2
other reductions follow: a) the sum of squares and b) degrees of freedom. Thus,
if after limiting the range, the error remains the same, $SS_{\text{err}}$ would
increase. Similarly, when the error is significantly reduced, $SS_{\text{err}}$
would increase. Therefore, any change in the error, should only be considered
significant if it is compensated by the amount of change in the degree of
freedom. For quantifying this, we defined two measures for linearity improvement
after limiting the data above $X_{min}$. The first measure, ``overal RMSE
change'' (oRMSE), was defined as:
 $$ oRMSE_i = {{{RMSE_n - RMSE_{n-i}} \over RMSE_n} *100}. $$
In parallel, ``relative RMSE change'' (rRMSE), was defined as:
$$ rRMSE_i = {{{RMSE_{n-i+1} - RMSE_{n-i}} \over RMSE_n} *100}. $$
, where $RMSE_n$ was the RMSE of the full length data. Next, these measures
were normalized to their maximum ($noRMSE$ and $nrRMSE$) and a 3rd
dimension was created by the distance of each pair ($noRMSE_i$, $nrRMSE_i$),
from the geometrical diagonal defined as $$ D = {det[(Q2-Q1) \cdot (P-Q1)]
\over \left \| (Q2-Q1) \right \|} $$, where P was the coordinates of a point
($noRMSE_i$,
$nrRMSE_i$) while Q1=[0 0] and Q2=[1 1] were the vertices of the geometrical
diagonal
of the RMSEs pair space. The point that had the maximum ``$(1-{D_i})$ +
$noRMSE_i$ + $nrRMSE_i$''  (this value can range between 0 to 3), was taken as
the optimal linearizing shortening index ($X_{min}$) (Fig.~\ref{fig13}D). Next,
we fitted all data ranges (from $N$ sample points to $N-X_{min}$) with the two
exponential models as described above.



\section*{Results}

In this study, we used data from multielectrode recordings in 3
species: cat motor cortex (cats i and ii with a 96 channel
multielectrode array in primary motor cortex, hindlimb area), cat parietal
cortex (cat iii, 8 bipolar electrodes), monkey motor cortex (three monkeys
with a 64 or 96 recordings from 96 channel multielectrode arrays in motor and/or
premotor cortex), and humans (2
patients with a 96 multielectrode array in middle temporal gyrus). In the
following, we briefly address definition of avalanche, then describe the results
of power-law analyses on spike avalanche, state-dependence, regional differences
and polarity-dependence of LFP maxima avalanche. At the end, we briefly
discuss alternative fits to the data.

\subsection*{Avalanche definition}

\begin{figure}[!h]
\section*{Figures}
\begin{center}
\includegraphics[width=3in]{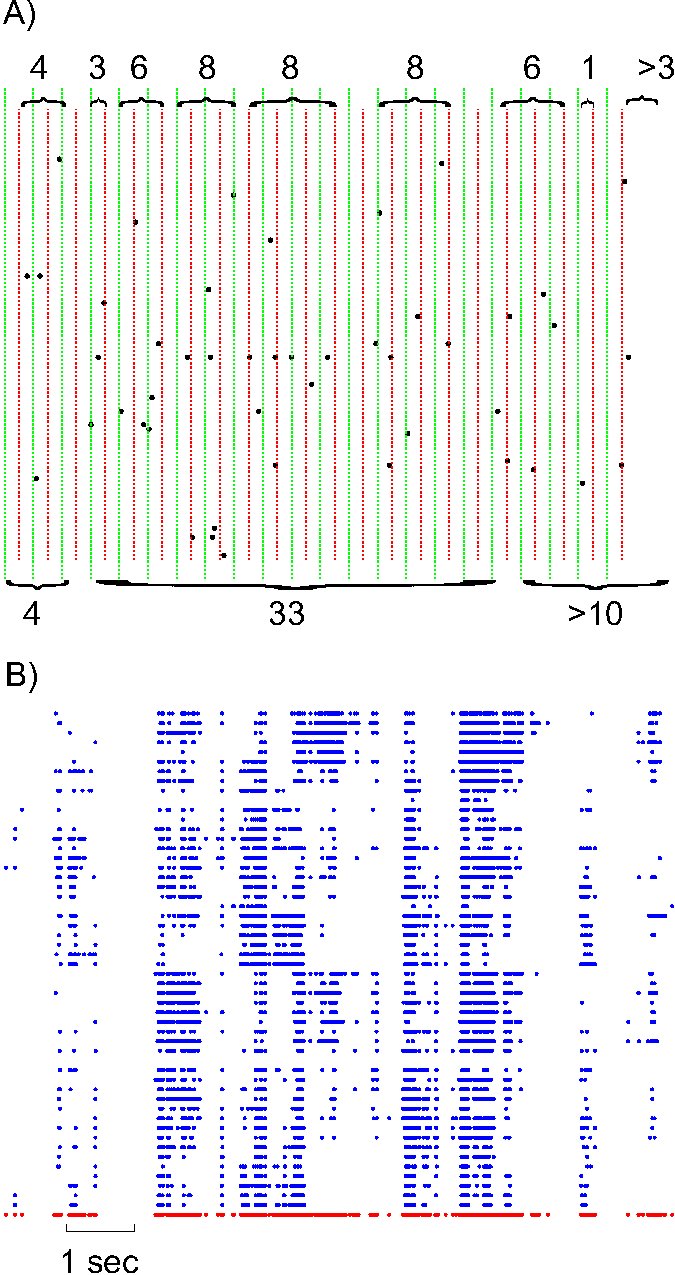}
\end{center}
\caption{ {\bf Definition of avalanches.}   {\bf (A)} comparison of
avalanche definition for 8ms vs 16 ms binning; green vertical lines
define the boundaries of 16ms binning; naturally, each 16ms bin is
composed of 2 independent 8ms bin (depicted with red dotted lines).
Accolades point to the avalanches, separated by quiescent periods. 
Top, 8ms avalanches and their sizes, Bottom: 16ms avalanches and
their corresponding size. Please note that last avalanche continues
after of the limits in this figure. {\bf (B)} negative local maxima
obtained from the grid of electrodes for a period of ~10 sec. Each
row represents negative maxima of a single LFP channel of a
selected threshold level $\geqslant$1.75$\times$STD of the
normalized LFP. The red dots in the bottom refer to ensemble
presence of nLFP maxima.}
\label{fig1}
\end{figure}

Figure~\ref{fig1} illustrates the definition of avalanche for
discrete (spike) and continuous (LFP) data, as they are used in
this study. For both spikes and LFP, we used bins of 1 to 15 ms
(in 2ms steps) for defining the quiescent vs active periods. 
Avalanches are defined by contiguous bins of non-zero activity,
separated by periods of quiescence (empty bins). The size of the
avalanche is defined as the sum of all activities (spikes or LFP
peaks) within that active period. Thus, the avalanches
depend on the bin size (as illustrated in Fig.~\ref{fig1}A for
spikes). For LFPs, we first discretized the continuous data based
on its local maxima. Both positive and negative maxima were
examined in our study. For each polarity, 17 levels of thresholds
were chosen (see Methods for details). After discretization, the
obtained matrix (Fig.~\ref{fig1}B) was used for the same binning
and avalanche definition as used for spike series.

\subsection*{Power-law fit}

It has been shown that that CDF provides a a better
measure than PDF as it avoids erroneous measures at the far end of
the distribution tail of probability curve (Newman, 2005; Clauset
et al., 2009). It is also necessary to exclude the values below the valid lower
bound, or else the calculated coefficient could be highly biased (Clauset et
al., 2009).
In each of the following estimates of power-law distribution, based on the
methods described previously, we
adapted the following steps on analyzing the CDF of avalanches: Values above
a given $X_{min}$ are used in a maximum likelihood estimate (MLE) of the
exponent $\alpha$. For each CDF, the proper lower bound of $X_{min}$ is selected
using a KS test. We also used 1000 semi-parametric repetitions of the fitting
procedure for obtaining estimates of uncertainty and goodness of fit.  

\subsection*{Avalanche analysis from spikes}

Next, we studied whether the spike avalanches follow power-law
distributions. 

\subsubsection*{Avalanche analysis in wakefulness} 

\begin{figure}[!ht]
\begin{center}
\includegraphics[width=5in]{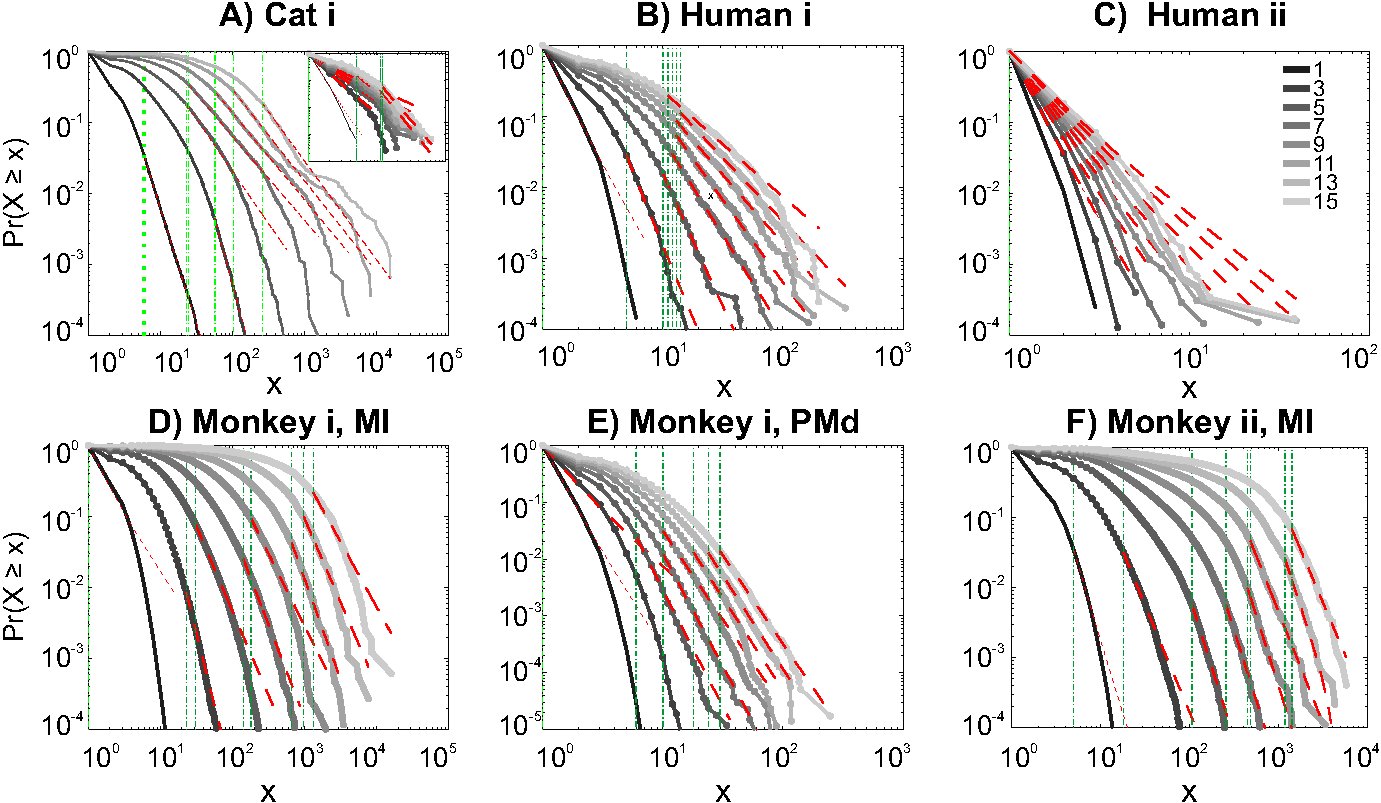}
\end{center}

\caption{{\bf Avalanche analysis on spiking activity during
wakefulness.} In idle awake ,{\bf (A)}. Cat i (96-electrode array)
and Cat iii (inset, 8-electrode array), {\bf (B)}. Human i
(96-electrode array), {\bf (C)}. Human ii (96-electrode array). 
Different line colors refer to different bin sizes as shown in the
legend. The lower bound ($X_{min}$, shown in green dotted line),
shows that the CDF of avalanche size, only partially, may follow
power-law distribution. Even in such cases, the exponents had very
high values, well above the criticality regime that is hypothesized
for 1/f noise. Panels {\bf (D)},{\bf (E)} and {\bf (F)}, show the same type of
curves for monkeys engaged in cognitive motor task (96-electrode array;
augmented with a 64-electrode array). Same pattern is observed; it
also seems MI has slightly higher values than PMd in the plausible
power-law regime. For the mean/std exponent values, see
Table.~\ref{tab1} and Table.~\ref{tab2}.}

\label{fig2}
\end{figure}

We first studied avalanche dynamics in awake resting recordings
from cats and humans. As depicted in Fig.~\ref{fig2}, neither of
these species, showed a dominant power-law behavior in their spike
avalanche size distribution. The average scaling exponent of awake
recordings for the decades that could be considered to follow
power-law (i.e.  $>$$X_{min}$), was to high to be related to SOC
systems (see Table.~\ref{tab1} and Table.~\ref{tab2} and
Fig.~\ref{fig2}.i,ii,iii). These values not only are distant from
those of 1/f noise, but also only apply to partial parts of the CDF
(cumulative distribution function) of avalanche sizes. These lack
of clear power-law characteristics is shown with $X_{min}$ lower
boundary (green dotted lines in Fig.~\ref{fig2}). Only values above
$X_{min}$ could "statistically" follow a power-law regime and as
mentioned, even in those cases, the exponent values were too high
to be considered a signature of SOC systems. It is important
to note that the CDF representation is cumulative, and thus the
left tail is not excluded from the data but its influence is
shifted to the right (see details in Clauset et al., 2009; see also
Methods).

\begin{figure}[!ht]
\begin{center}
\includegraphics[width=4in]{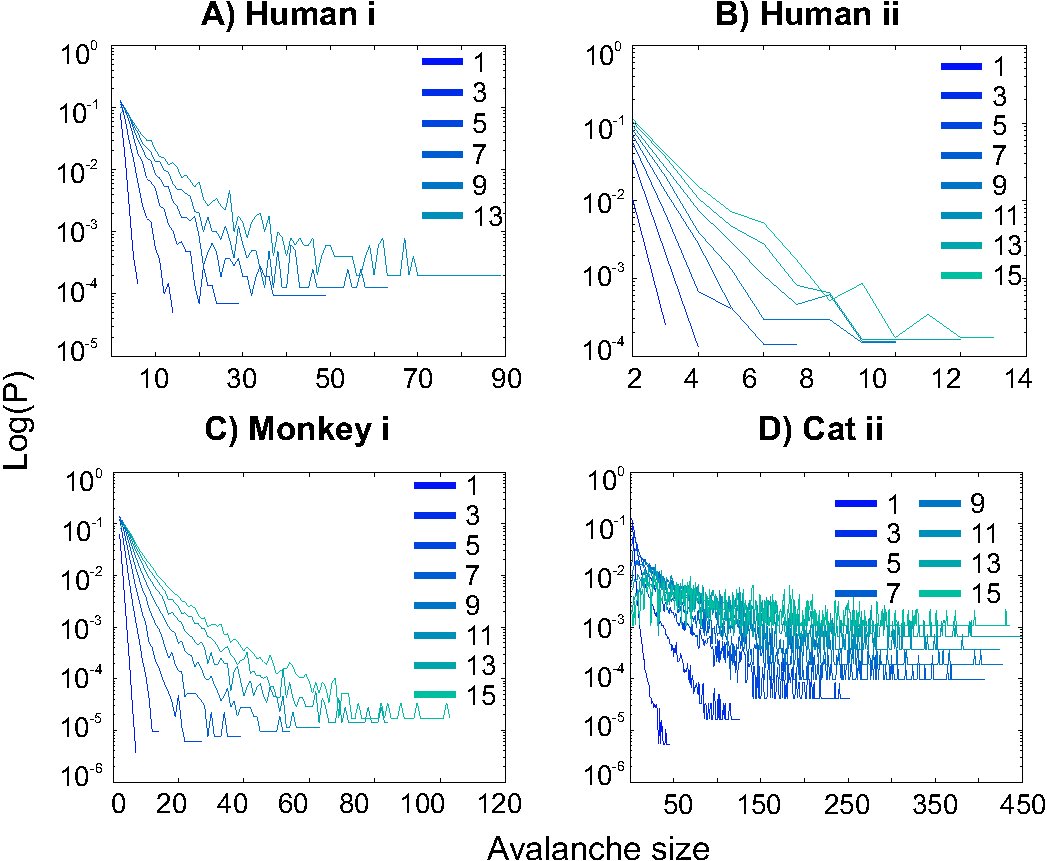}
\end{center}
\caption{ {\bf Spike avalanche distributions in log-linear
representation.} Different line colors refer to different bin sizes
as shown in the legend. An exponential process has a linear trend
in log-linear scale. Spike avalanches for all coarse graining
levels, showed a linear trend. Please notice that bin sizes 11 and
15 are not shown because for the clarity in the line plot, but
showed similar linear trend in this scale (not shown).}
\label{fig3}
\end{figure}

Interestingly, representing the size distributions in log-linear
scale revealed a scaling very close to linear for all species
(Fig.~\ref{fig3}), indicating that avalanches defined from spikes
scale close to an exponential, as would be predicted by a
Poisson-type stochastic process. This conclusion was also reached
previously by analyzing units and LFP recordings in cats (Bedard et
al., 2006). Also, as can be seen in the inset of panel A of
Fig.~\ref{fig2}, the same analyses done on the awake recording from
the parietal cortex (albeit spatially sampled at only 8 electrodes)
shows similar scaling behavior.

In addition to wake resting recordings, we also considered
recordings made while monkeys engaged in cognitive motor tasks. 
Similar to awake resting recordings in cat and man, the lower bound
was variable between different binning sizes, thus excluding parts
of the "invalid" initial avalanche sizes, which are usually used as
evidence of existence of power-law (Beggs and Plenz, 2003;
Petermann et al., 2009; Klaus et al., 2011). The inclusion of these
initial parts may cause errors, and were removed here;
however, their cumulative effects are still present in the tested
regimen above $X_{min}$ of the analyzed ``cumulative distribution
function'' (Newman, 2005; Clauset et al., 2009; Bauke, 2007;
Goldstein et al., 2004). Above the lower bound value, all the CDF
curves showed significant high exponent values. Interestingly, the
MI (in both monkeys A and B) had similar mean to PMd
(Table.~\ref{tab1}, Fig.~\ref{fig2}D,E,F), suggesting similar
dynamics in the two areas.

\subsubsection*{Avalanche analysis during natural sleep}

\begin{figure}[!ht]
\begin{center}
\includegraphics[width=4in]{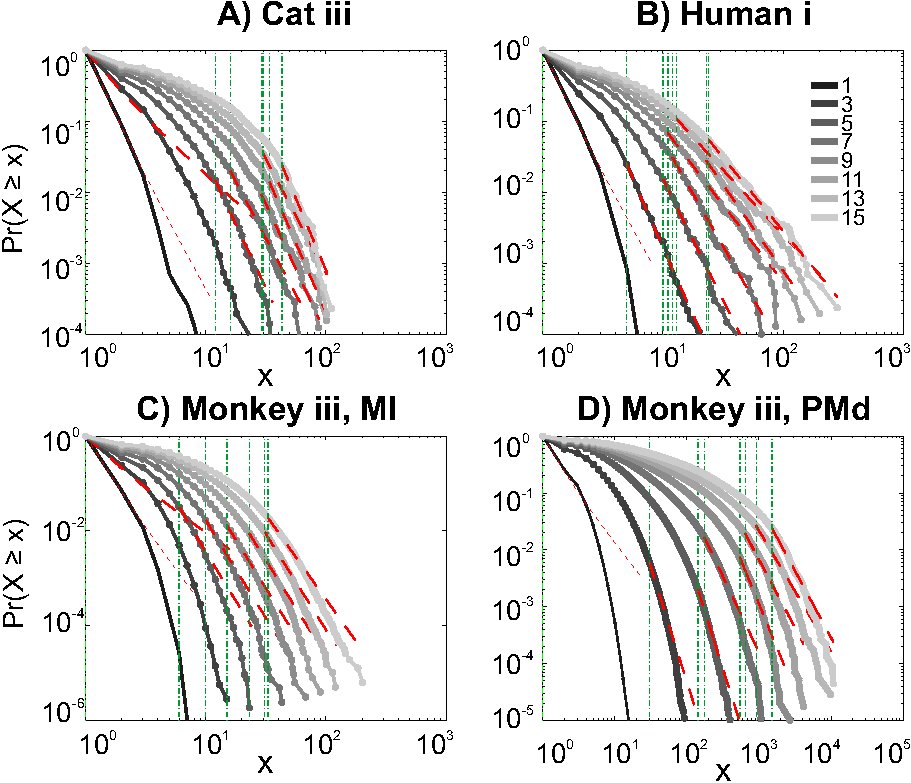}
\end{center}

\caption{{\bf Avalanche analysis of spiking activity during
slow-wave sleep.}  {\bf (A)} Cat iii, {\bf (B)} Human i, {\bf (C)} monkey iii MI
and {\bf (D)} monkey iii PMd. Different line colors refer to different
bin sizes as shown in the legend.  In parallel to awake dynamics
(Figure 2), there is no sign of criticality, the curves follow
different partial power-law with high exponents and variable lower
bound values. The avalanche dynamics do not show a state-dependent
trend.  For the mean/std exponent values, see Table.~\ref{tab1}.}

\label{fig4}
\end{figure}

\begin{figure}[!ht]
\begin{center}
\includegraphics[width=4in]{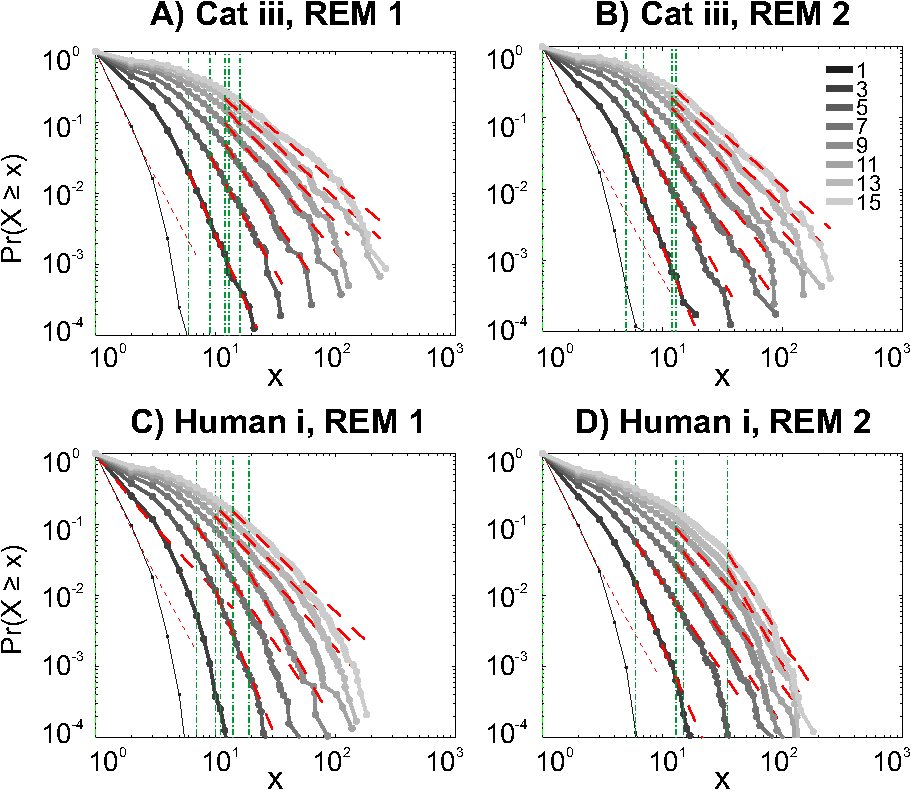}
\end{center}

\caption{{\bf Avalanche analysis of spiking activity during REM
sleep.} {\bf (A)} cat iii REM episode 1, {\bf (B)} cat iii REM episode 2, {\bf
(C)} human i REM episode 1, {\bf (D)} human i REM episode 2. Different line
colors
refer to different bin sizes as shown in the legend. Similar to
awake and SWS, the lack of criticality, variability through
different coarse graining thresholds and lower bounds is the
universal finding.  For the mean/std exponent values, see
Table.~\ref{tab1}.}

\label{fig5}
\end{figure}

It has been claimed that wakefulness may not be the best state to
display SOC, and that avalanches may be more naturally related to
brain states with oscillations, and slow-wave oscillations in
particular (Hahn et al., 2011). In contrast to this, a previous
study in cat found that like wakefulness, slow-wave sleep (SWS) did
not display power-law scaling as defined from spike avalanches
(Bedard et al., 2006), but this latter study suffered from a
limited spatial sampling. To further investigate the issue, we
have examined SWS and Rapid Eye Movement (REM) sleep periods with
more dense sampling of spike activity. Figures~\ref{fig4} and
\ref{fig5}, show the analyses for cat, human i and ii as well as
monkey iii (MI and PMd) for SWS and REM periods.  In none of these
cases we see clear sign of power-law scaling. In all cases (except
human ii), the variability of lower bound between different bin
sizes is robust. All the curves represent "partiality of
power-law" with high exponent values. During SWS, cat, human
subjects and monkey iii (MI and PMd) all manifested either lack of
significant power-law scaling, or had such higher exponent values
that makes it highly unlikely for power-law to be the generating
process of spike dynamics (Table.~\ref{tab1}). Similarly, in REM
periods, there was no evidence for power-law scaling in human i's
first and second REM episodes. Together, with Cat REMs' high
exponents values, power-law scaling appears to be an unlikely
candidate to describe the statistics of neural firing
(Table.~\ref{tab1}). Taken together, these various tests all based
on proper statistical inferences, show that spike avalanches do not
follow power-law scaling, for any brain state or sampling density.

Detailed numerical values for spike avalanche CDF exponents and
their goodness of fit are provided in Table.~\ref{tab1} and
Table.~\ref{tab2}.  

\subsection*{Avalanche dynamics from local field potentials}

Next, we investigated the occurrence of avalanche type of dynamics
from the local field potentials, which were simultaneously recorded
with unit activity, in all datasets.

\subsubsection*{Relation between LFP peaks and spiking activity} 

\begin{figure}[!ht]
\begin{center}
\includegraphics[width=3in]{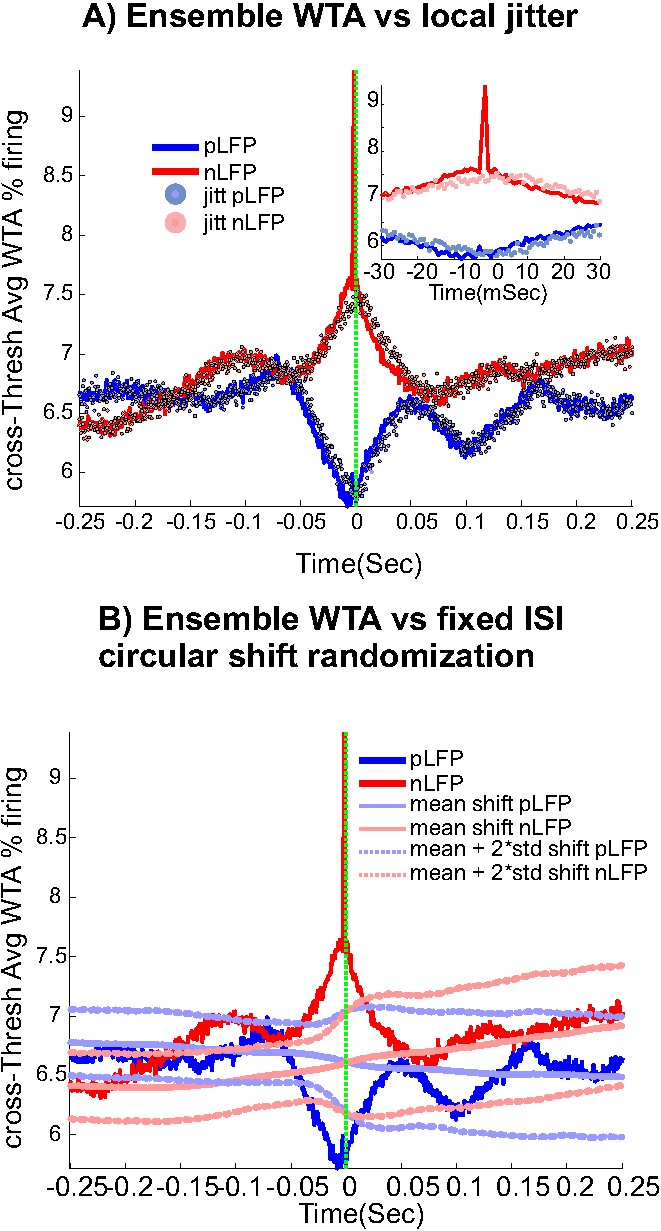}
\end{center}
\caption{{\bf Relation between unit firing and LFP peaks in
wakefulness.} nLFP (red) and pLFP(blue)-based wave-triggered
average (WTA) of percentage unit activity, showing that the negative
peaks have closer association with an increase of neuronal
firing. {\bf (A)} Tightly regulated local jitter of nLFP peaks destroys
the large nLFP peak. Inset shows the zoom around 0. {\bf (B)} Preserving the
internal structure of aggregate spike train and ensemble LFP peaks, but
destroying the relation between the two leads to the disappearance of the nLFP
peak. See text for details of randomization and controls. The WTA traces in
this figure are from Human i, (based on 183127, 98520 and 47451 nLFP and
158737, 79225 and 36020 pLFP peaks for low, medium and high threshold
respectively.) 
}
\label{fig6}
\end{figure}

Calculation of neuronal avalanches from LFP data is based on the
assumption that statistically speaking, in comparison with
the positive LFPs (pLFP), the negative LFP (nLFP) peaks are
more strongly related to neuronal activity (e.g., see
Destexhe et al., 1999 and references therein). Indeed, the
8-electrode cat LFP data analyzed here show such a relation
(Destexhe et al., 1999; Touboul and Destexhe, 2010). To further
test this relation, we also examined the simultaneous LFP and unit
recordings in the ensemble recordings in cat, man and
monkey. We used a wave-triggered-average (WTA) procedure, where
the ensemble of nLFPs were used to epoch the ensemble spike
activity. Averaging across these WTAs across different thresholds,
show that there is indeed a weak relationship between nLFP and
spiking (Fig.~\ref{fig6}A). However, repeating the same procedure
for positive LFP (LFP) peaks, did not display any relation
(Fig.~\ref{fig6}B), in agreement with the same analysis in cats
(Touboul and Destexhe, 2010). Through four different types of control
and randomization, we show that the relation between nLFP and spike is robust
and is not attributable to randomness of the spiking events or spurious
fluctuations in the LFPs. For details of these control/randomization, see
methods and Fig.~\ref{fig6}. This fundamental difference between
nLFP and pLFP peaks provides a very important test to infer if a
given power-law observation from LFPs is related to the underlying
neuronal activity, as we will see below.

\subsubsection*{nLFP avalanches}

\begin{figure}[!ht]
\begin{center}
\includegraphics[width=5in]{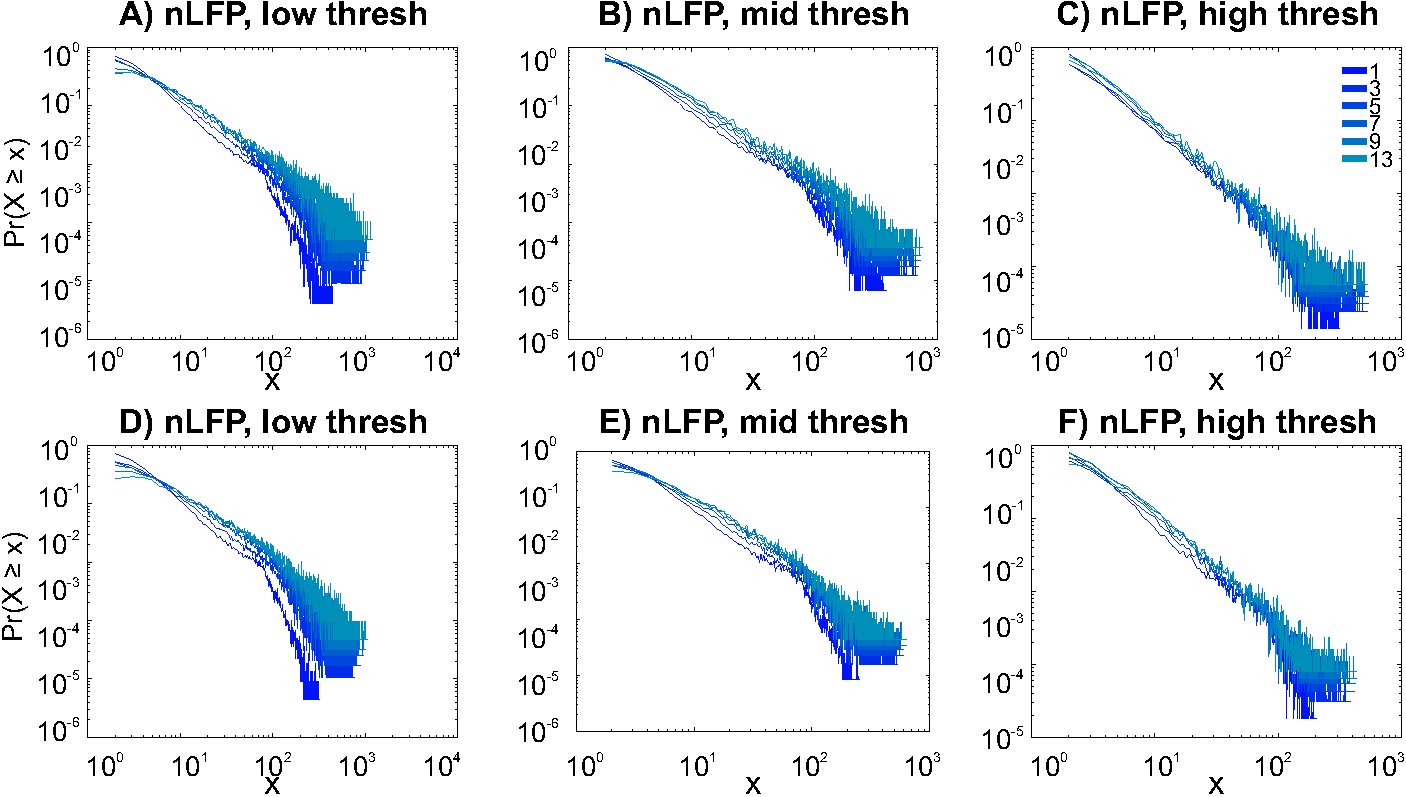}
\end{center}
\caption{ {\bf Avalanche analysis in awake monkey LFPs in
logarithmic representation.}  A power-law process has a linear trend in
log-log scale. LFP (negative or positive) maxima avalanches for all coarse
graining levels, as well as all thresholds, showed a linear trend. Please
notice that bin sizes 11 and 15 are not shown because for the clarity in the
line plot; however, they too, also showed a very clear linear trend in this
scale. Such trend is necessary but not sufficient for a process to be
power-law.
See text and Fig.~\ref{fig9}}
\label{fig7}
\end{figure}

\begin{figure}[!ht]
\begin{center}
\includegraphics[width=5in]{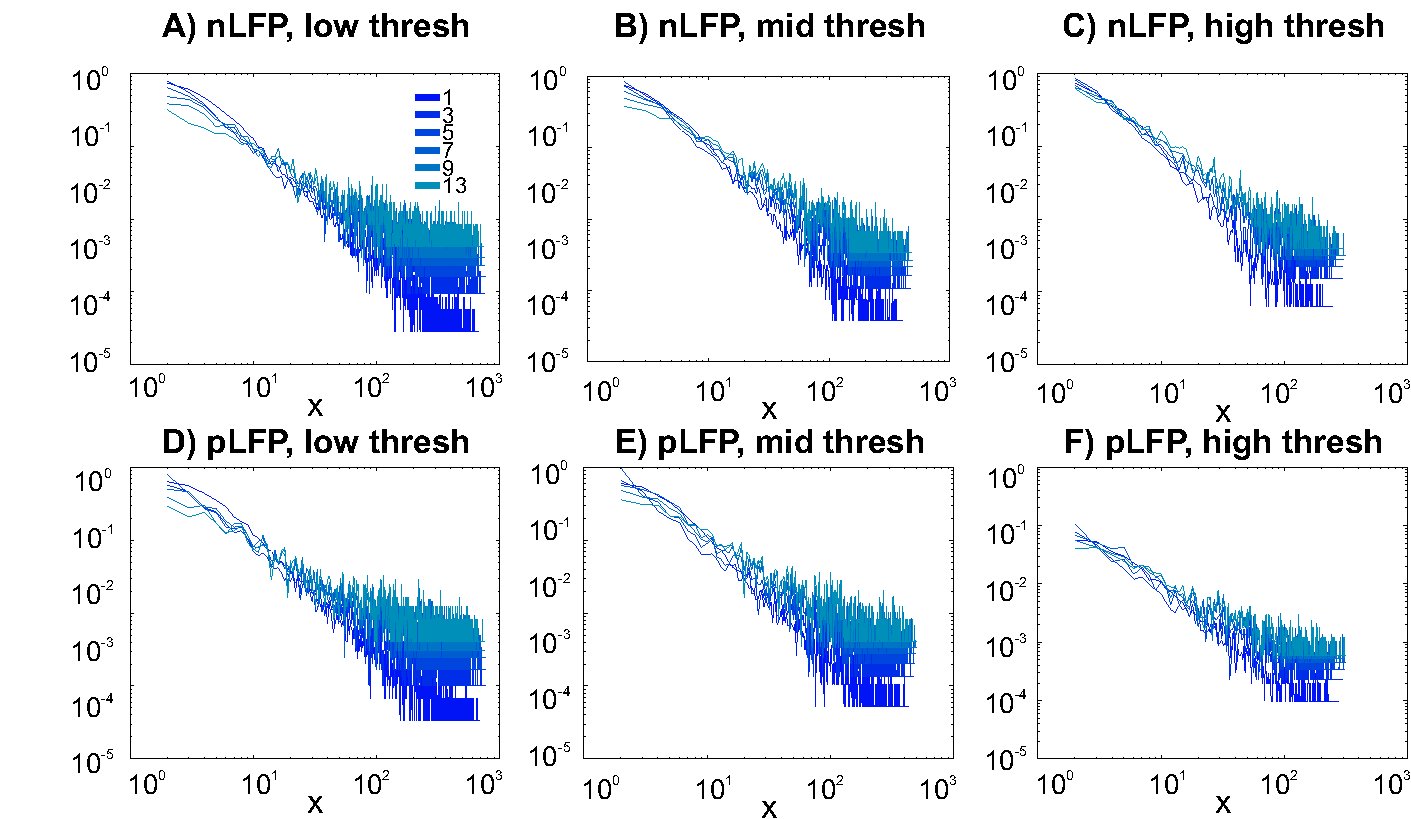}
\end{center}
\caption{ {\bf Avalanche analysis in awake human LFP in logarithmic
representation.}   A power-law process has a linear trend in log-log
scale. LFP (negative or positive) maxima avalanches for all coarse graining
levels,  as well as all thresholds, showed
a linear trend. Please
notice that bin sizes 11 and 15 are not shown because for the clarity in the
line plot; however, they too, also showed a very clear linear trend in this
scale. Such trend is necessary but not sufficient for a process to be
power-law.
See text and Fig.~\ref{fig9}}
\label{fig8}
\end{figure}

\begin{figure}[!ht]
\begin{center}
\includegraphics[width=4in]{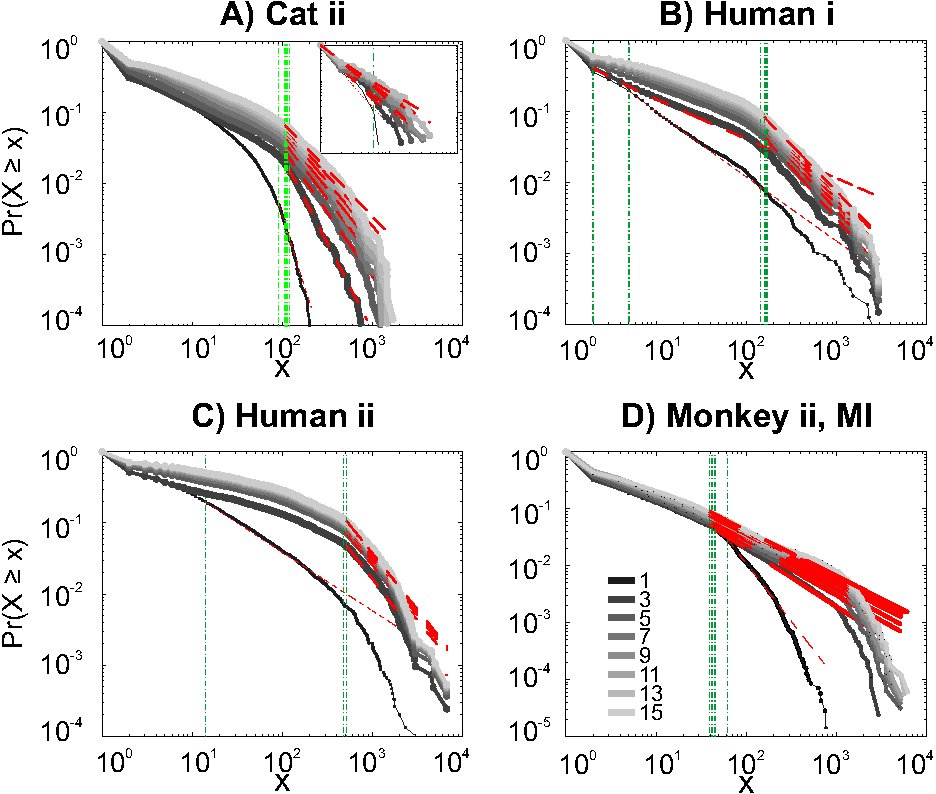}
\end{center}

\caption{{\bf Avalanche analysis based on LFP negative peaks in
wakefulness.} {\bf (A)} Cat ii (96 electrode array) and Cat iii (inset,
8 electrode array), {\bf (B)} Human i, {\bf (C)} Human ii , {\bf (D)} Monkey ii
MI. In all cases, different binnings lead to variable lower bound and
scaling exponents. Lack of linear trend in CDF shows that the
observed linear trend in log-log scale, as shown in Fig.~\ref{fig7}
and Fig.~\ref{fig8}, are not sufficient for showing that avalanche
dynamics are power-law processes. For the mean/std exponent values,
see Table.~\ref{tab3}.}

\label{fig9}
\end{figure}

Similar to previous studies, we investigated the avalanche dynamics
from nLFPs. The nLFPs were detected using a fixed threshold,
defined as a multiple of the standard deviation (STD) of the LFP
signal (see Methods), and several thresholds were tested.  In the
following, we use ``high'', ``medium'' and ``low'' thresholds,
which correspond to 2.25, 1.75 and 1.25 multiples of the standard
deviation, respectively. As shown in Figs.~\ref{fig7} and
\ref{fig8}, the distributions defined for avalanches at different
bin sizes and thresholds seem to display power-law scaling, both
for human and monkey. This result seems to be in agreement with
similar analyses done on awake monkey (Petermann et al., 2009). 
However, plotting the same data as CDF revealed that the scaling as
power-law was very narrow (Fig.~\ref{fig9}). While Monkey ii
displayed apparent power-law over more than one decade, the other
cases from cats and humans, did not display any convincing
power-law scaling. For details of nLFP avalanches for an example
subject, and its comparison with pLFP avalanches, see
Table.~\ref{tab3}. One can also note that in some of the CDFs
(and their counterpart PDF), there is a possibility that the
distribution can be segmented into two regions each covering
certain decades of avalanche size. In such cases, relying on a single
scaling exponent to describe the totality of the functional
dynamics of the network does not seem adequate. This could be an
indication that the space of the distributions is not uniform and
the underlying mechanisms could be of metastability nature 
(Mastromatteo and Marsili, 2011). In such scenario, interaction
with the external world could push the system from the ``currently
most stable state'' to a new ``most stable state''. Such constant
changes may lead to the formation of nonuniform distribution of the
neural events at different temporal scales. Therefore it is
essential to emphasize that, in some cases, one scaling exponent
may not be sufficient to describe the complexity of the spiking or
oscillations.

\subsubsection*{pLFP avalanches}

\begin{figure}[!ht]
\begin{center}
\includegraphics[width=5in]{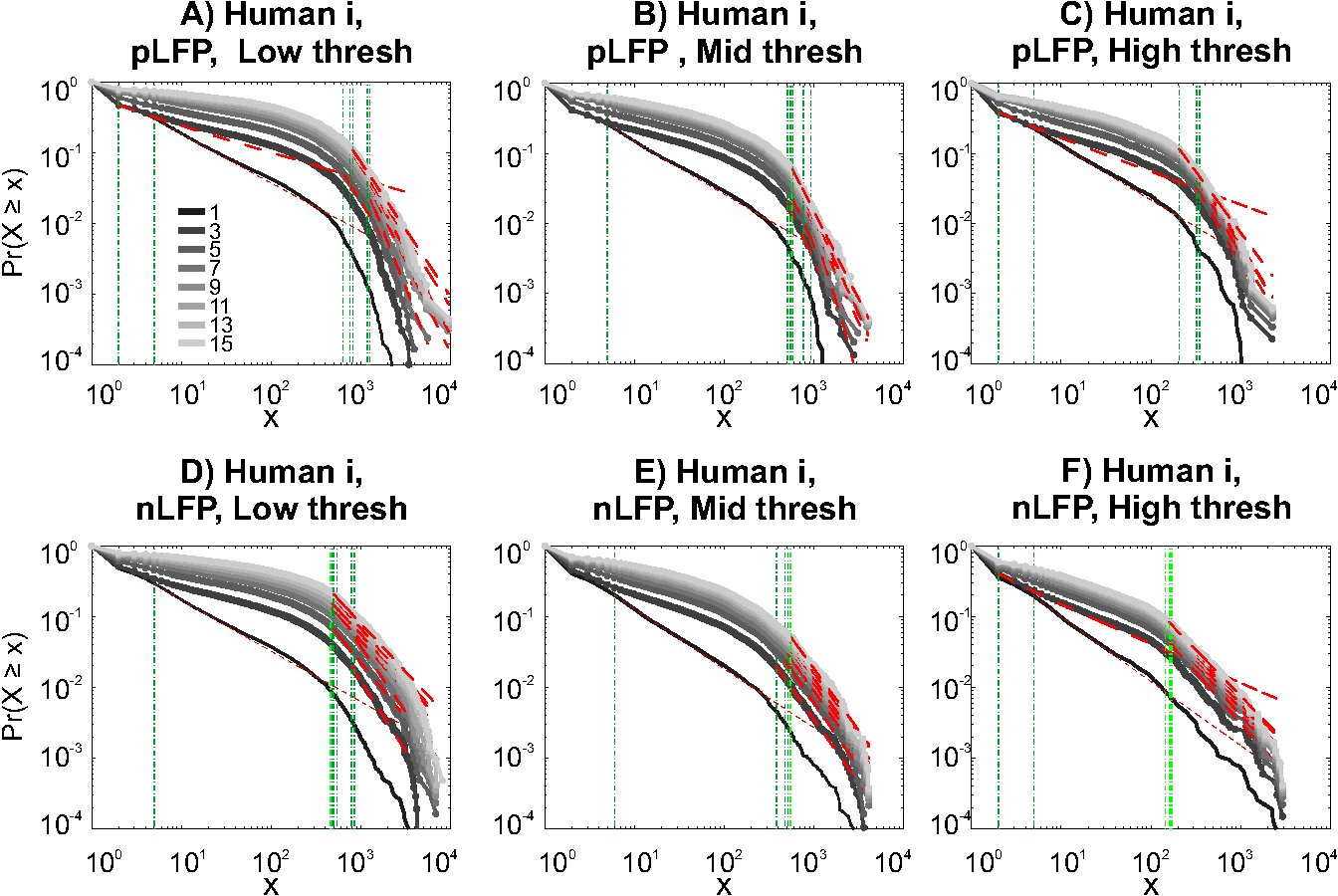}
\end{center}
\caption{{\bf Comparison of Avalanche analysis based on negative and positive
peaks.} LFP (negative or positive) maxima avalanches for all coarse graining
levels, as well as all thresholds did not show linear trend in CDF, therefore
negate power-law as the generating process. These curves show while nLFP has a
closer relation
with spiking, the avalanche dynamics of nLFP and pLFP are
strikingly similar in their lack of robust criticality when tested
with rigorous statistical tests.}
\label{fig10}
\end{figure}

Next, we investigated the avalanche dynamics of positive LFP peaks,
which, as we have seen above, is not statistically related to
firing activity (Fig.~\ref{fig6}). Similar to nLFP peaks, the pLFP
avalanches defined for human wakefulness did not display power-law
scaling (Fig.~\ref{fig10}). Both nLFP and pLFP had similar CDF of
avalanche size across different species and cortices. The example
shown in Fig.~\ref{fig10} (awake human) shows that across different
thresholds, both nLFP and pLFP had variable lower bounds and high
scaling exponents for the region of the data that could
statistically be considered for power-law properties. Moreover,
the absence of any region with clear linear scaling in the
logarithmic coordinates further confirms that there is no power-law
scaling in this case. For details, see Table.~\ref{tab3}.

\subsubsection*{Avalanches in different cortical regions}

\begin{figure}[!ht]
\begin{center}
\includegraphics[width=5in]{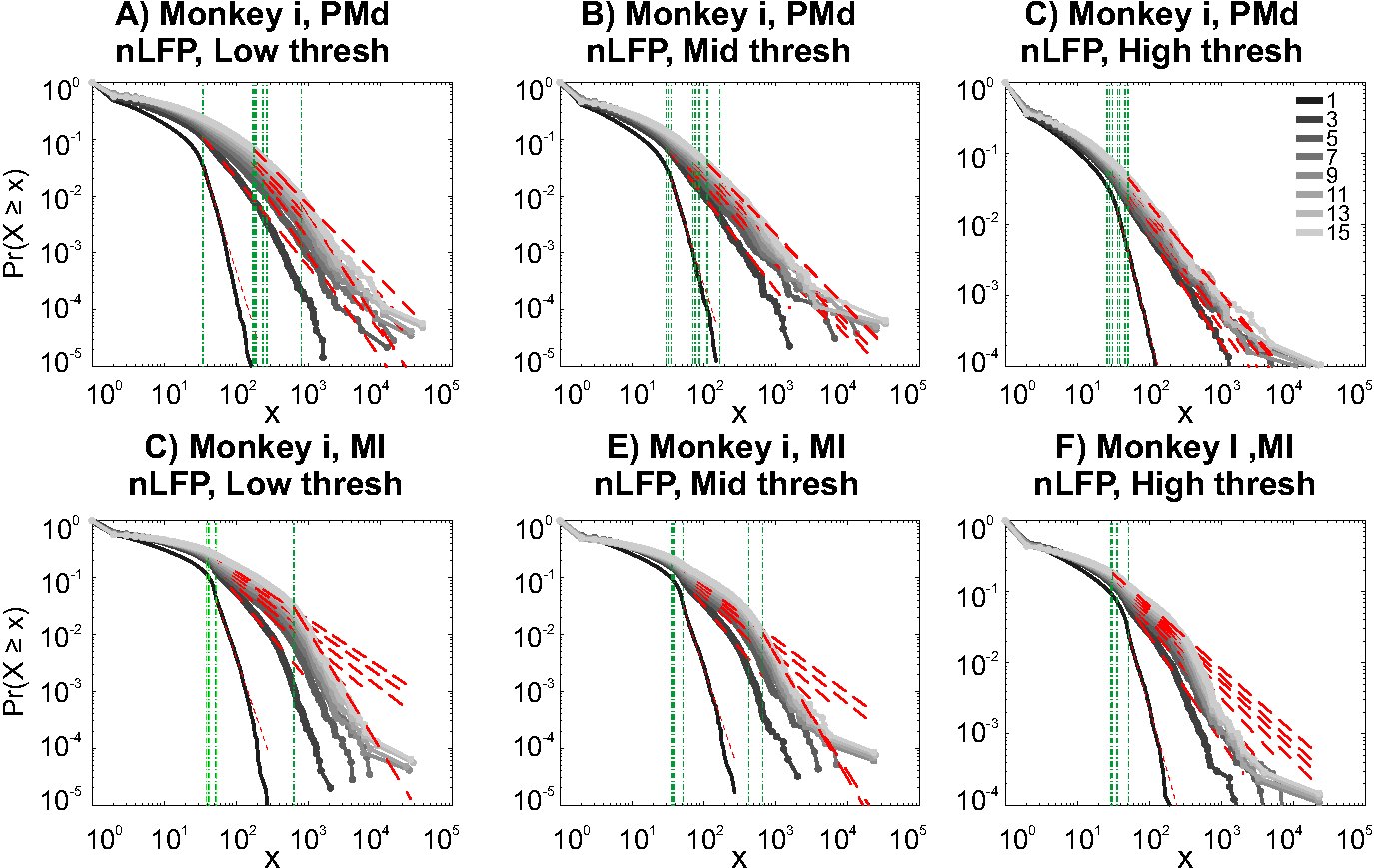}
\end{center}
\caption{{\bf Avalanche analysis in different cortical areas
recorded simultaneously.} Avalanche dynamics in nLFP shows that the
CDF of the input and output units of two interacting cortices have
slightly different characteristics but neither follow criticality
regime. {\bf (A)} Monkey i, MI, low threshold {\bf (B)} Monkey i, MI, medium
threshold, {\bf (C)} Monkey i, MI, high threshold, {\bf (D)} Monkey i, PMd, low
threshold {\bf (E)} Monkey i, PMd, medium threshold, {\bf (F)} Monkey i, PMd,
high
threshold.}
\label{fig11}
\end{figure}

In the cases that we had simultaneous, dual array multielectrode
recordings from PMd and MI, the analyses showed that these two
cortical areas do not show signs of criticality but have slight
differences in their exponent values for MI and for PMd
(Table.~\ref{tab1} and Table.~\ref{tab2}, Fig.~\ref{fig11}). Such
findings show that the fact that these two cortices directly
interact with each other, and one acts as input and one as the
output of motor processing unit, is reflected in their slightly
different CDF features. Thus, two different cortical areas seem to
display similar features, although no sign of power-law scaling.

\subsection*{Statistical analysis of the avalanche distributions}

\subsection*{Goodness of fit}

Given data $x$ and given lower cutoff for the power-law behavior
$X_{min}$, we computed the corresponding p-value for the
Kolmogorov-Smirnov test, according to the method described in
Clauset et al.\ (2009). See methods for details. The results are
given in Tables~\ref{tab1}, ~\ref{tab2} and ~\ref{tab3} (``gof''
columns).

\subsection*{Avalanche size boundaries}

{Imposing lower or upper bounds when fitting avalanche
distributions can greatly affect the outcome of the fit (Clauset et
al., 2009).  In many cases, the analyses have been limited to the
lower boundary of avalanche size $=$ 1 and $X_{max}$ of N, where N
is the number of channels.  Using such bounds improves the fitting
of the data by power-law compared to other distributions, as
confirmed by KS-statistics (Klaus et al 2011). The pitfalls of such
an approach are two-fold: a) the lower boundary is set to 1,
therefore the avalanches that are below the acceptable lower bound
of $X_{min}$ are erroneously fitted with the power-law, thus
reducing the reliability of the fit while producing mis-estimated
scaling exponents (see Clauset et al., 2009 for details of lower
bound selection). b) $X_{max}$ is set to the maximum active
channels, and any return to a given channel is counted in the
avalanche, but the maximum allowed avalanche size is limited to N,
based on the argument that the large avalanches are infrequent and
their inclusion implies misfit. This type of approach, limits the
number of avalanches to an extreme degree and introduces a bias. 
Below we investigate this bias.}

\subsubsection*{Avalanche size distribution and upper boundary limits}

\begin{figure}[!ht]
\begin{center}
\includegraphics[width=4in]{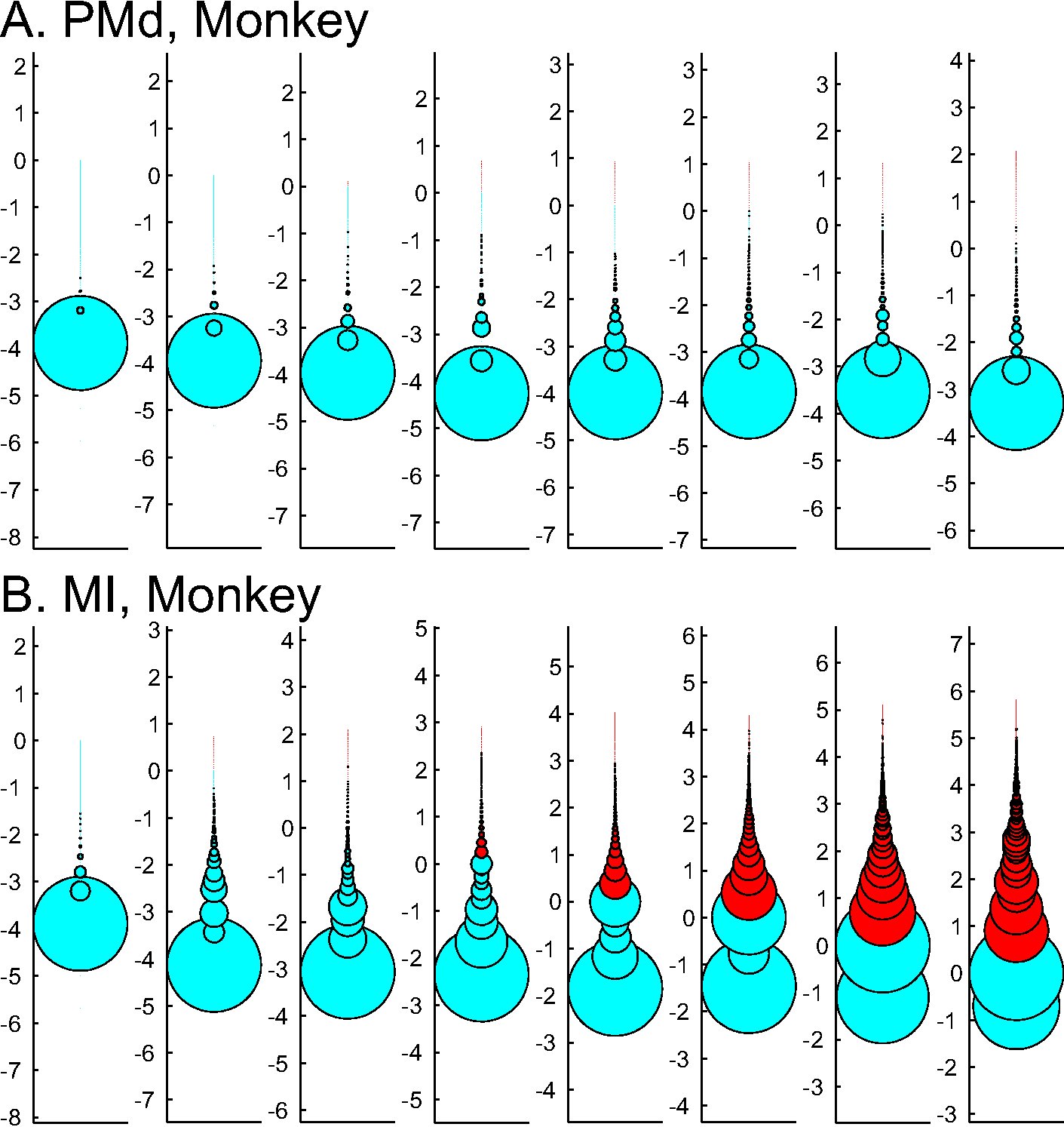}
\end{center}

\caption{{{\bf Effects of setting upper boundaries on
avalanche size distribution.}} {Each column shows avalanches
of a different bin size (increasing from left to right).Panel {\bf
(A)} and {\bf (B)}, show the results of spike avalanche size
distribution of the PMd and MI (respectively). For each bin size,
the distributions of different avalanche sizes are shown in
circles; the avalanche size increases from the bottom to the top,
while the size of each circle represents the ratio to the overall
number of avalanches. Red color shows the excluded (saturated)
avalanches enforced by limiting the Xmax to N (number of
independent measures; i.e. units in the case of spike avalanches
and electrodes in the case of LFP avalanches).  Cyan color shows
the included avalanches. Y axis is in logarithmic scale for better
visualization and the values of Y represent the orders of magnitude
of N for proper comparison between different bin sizes (i.e. a
given circle at y=2, represents the avalanches that their size
=2log(N), its diameter shows the number of avalanches that had that
size and its color shows whether it is included or excluded
according to the Xmax=N rule).}}

\label{fig12}
\end{figure}

{Fig.~\ref{fig12} tests the effect of enforcing an upper
boundary to the avalanche analysis. The red color shows the
excluded (saturated) avalanches enforced by limiting the $X_{max}$
to N (number of independent measures), while cyan represents the
acceptable avalanches below this upper threshold.  This figure
shows that setting the $X_{max}$ to a cutoff value of N, produces
variable biases based on the bin size.  Importantly, in
simultaneously recorded regions, the majority of avalanches will be
included in one case (like in PMd as shown in panel A) but not in
the other (like MI, as depicted in panel B).  Such discrepancy
emphasizes that setting a cutoff will necessarily introduce a bias
and causes variable results from region to region and from bin size
to bin size.}

\subsubsection*{Comparison of exponential and power-law fit: Model
Mis-specification and lower boundary problem}

{It has been argued whether neuronal avalanches are better
fitted by an exponential or power-law distribution. Here we tested
two aspects, exponential vs.\ power-law comparison, as well as the
effect of setting a lower boundary to the fit. It has been shown
that defining a proper lower boundary improves the maximum
likelihood that the distribution could be fit by a power-law
(Clauset et al., 2009).  In agreement with this, Klaus et al. 
(2011) used a lower boundary of 1 and showed that using
KS-statistics, the power-law indeed provides a better fit to the
data in comparison to exponential distribution. Here, we
systematically tested whether such practice would return erroneous
results in avalanche analysis. The results shown in
Fig.~\ref{fig13}A,B, are from cat spikes data.  For each bin size,
we first defined the optimal lower boundary after Clauset et al. 
(2009; see Methods), called $X_{min}$.  We started with a lower
boundary set to 1, and reduced the distribution of avalanche data
gradually up to $X_{min}$. For each newly produced set, we
calculated the empirical CDF (ECDF) as well as the provisional
fitted probability's CDF (based on direct maximum likelihood) for
both exponential as well as power-law.  The results for a sample
bin size are shown in Fig.~\ref{fig13}A.  Power-law at the lower
boundary of 1 provides a bad fit. However, overall, power-law
outperforms the exponential fit, specially after limiting the range
of the data by increasing the lower boundary.  The best power-law
fit is obtained when the lower boundary approaches $X_{min}$.}

{This finding matches the results of the KS test (based on
Clauset et al., 2009) as we report in this manuscript. However,
from our analyses, we know that when we reach the best power-law
fit, the estimated scaling exponents are too high for any known
natural system to follow a self-organized criticality regime.
Therefore, we have a situation where either one gets unreliable but
desired scaling exponent by setting the lower boundary to 1, or one
obtains reliable but undesired scaling exponent by setting the
lower boundary to $X_{min}$>1.}

{Next, we quantified the goodness of fit with a more rigorous
approach than the simple KS test. If the parametric CDF is close to
the probabilities from the ECDF, then the depicted line should
approach the diagonal (1:1) line with minimal drift from it.  For
quantifying this, we measured the integral of the distance of each
point on the p-p curves from the 1:1 diagonal line.  This value
should be zero in a perfect fit; its non-zero value shows departure
from a perfect fit.  Fig.~\ref{fig13}B shows the results for all
bin sizes.  Similar to Klaus et al.\ (2011), the power-law provides
a better fit in comparison to exponential.  However, there are two
aspects that can not be ignored for this condition to be true: a)
the distance improves only as we tighten the lower bound criteria
to be close to $X_{min}$, but it does not mean that this is a
proper fit.; b) there is no rule of thumb
for such an improvement; in almost all of the cases, a linear
relationship in the normal probability plot distribution of the
distance was not found. This shows that
power-law provides a better fit than the exponential distribution, 
but that both fits are not satisfactory.  We consider alternative 
distributions below.}

\subsection*{Alternative distributions for avalanche dynamics}

Although previously, at the microcircuit scale, some studies have
asserted the existence of criticality as a universal characteristic
of neural dynamics in both spike and LFP avalanches (Beggs and
Plenz, 2003; Ribeiro et al., 2010), other evidence suggest that
same behavior can also bee observed through stochastic processes
(Bedard et al., 2006; Touboul and Destexhe, 2010). In this study,
after rigorous testing, we showed that the avalanches do not follow
power-law as a universal feature. Thus we also tested whether an
alternative probability distribution could provide a better
estimate for the experimental observations.  

\begin{figure}[!ht]
\begin{center}
\includegraphics[width=3in]{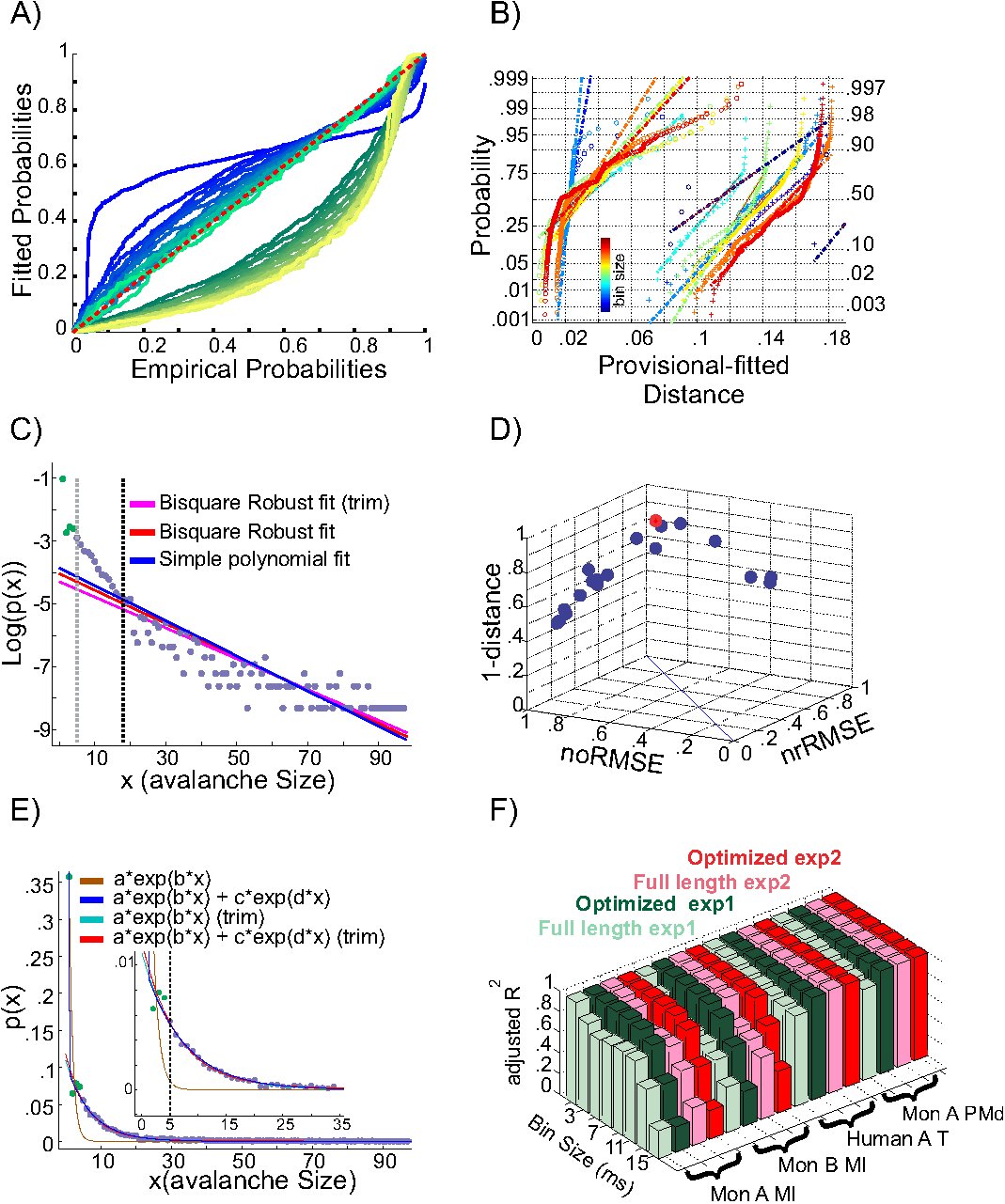}
\end{center}

\caption{{\bf A-B: Fits comparison and lower boundary.}{\bf
C-F:Alternative fits for avalanche size distributions.} {
(A) Probability-Probability plot (ECDF vs provisional CDF) for a
sample bin-size (cat i spike avalanche). Green colors are p-p plot
for ECDF vs exponential, and blue colors are for p-p plot for ECDF
vs power-law. In each color family, as the lower boundary is
increased (from 1 to Xmin), the color saturation fades; i.e. 
darkest color shows lower boundary of 1 and the lightest shows
lower boundary of Xmin (where Xmin is based on the Clauset method
for fitting power-law to empirical data). {\bf (B)} Integral of
p-p distance to the 1:1 diagonal (perfect match of the parametric
CDF to ECDF). The colors (blue to red) are related to bin sizes
(from smallest to biggest). Cross signs represents exponential
distance and circles represents power law distance to the ECDF.}
{\bf (C)} Simple exponential fitting of spike avalanche data.  The
data points (purple and green) are plotted in a log-linear
representation, together with a simple polynomial fit (blue), a
robust fit calculated on the full length data (red) and a robust
fit on the reduced data (magenta).  The two vertical lines indicate
the lower bound of the region of linearity, i.e.  ``$X_{min}$'',
calculated based on the simple polynomial fit (black) and the
bi-square method (grey).  {\bf (D)} Comparison of the goodness of
fit of different exponential fits to different reductions of the
same dataset.  The 3 coordinates are ``normalized overall
improvement of RMSE'' (noRMSE), ``normalized relative improvement
of RMSE'' (nrRMSE) and distance of a point from the diagonal in
(noRMSE,nrRMSE) plane.  Each point in this 3D space, is the result
of a bi-square robust fit after elimination of the first $\it{i}$
elements of the data (best fit in red).  {\bf (E)} Bi-exponential
fitting of the same data.  The ``sum of exponential'' model (exp2)
gave a very good performance in both full length (dark blue) and
reduced above ``$X_{min}$'' (red).  The ``simple exponential''
model (exp1) reaches a very good fit only for the reduced set
(cyan) but not for the full length of the avalanches (light brown).
{\bf (F)} Effects of linearity improvement on exponential fits. 
Each set of four colors refer to the spike avalanche of Monkey i
(MI), Monkey ii (MI), Human A(Temporal) and Monkey i (PMd). In each
set, green colors refer to the simple exponential family (exp1) and
the red colors depict the sum of exponentials (exp2). Light green
and light red, refer to the calculated $\bar R^2$ on full length
avalanche sizes, while dark green and red show the average $\bar
R^2$ for the dataset ranging from $\it{N-1}$ to $ \it{N -
{X_{min}}}$ where the optimized length $X_{min}$ was 5 (see panels
C and D). Panels C, D and E were obtained from 15 ms bin avalanches
from human i awake spikes.}

\label{fig13}
\end{figure}

We first tested a simple exponential fitting of the spike
avalanches, by fitting straight lines in a log-linear plot.  As
seen from Fig.~\ref{fig13}C, a linear fit (``exp1'') can only fit
part of the data, as the initial points (for small size) do not
scale linearly. In detection of the lower bound of linearity, i.e.
($X_{min}$), the robust bi-square method is more stringent than
simple least square fits and leaves behind more data points for
exponential fitting (see different lines in Fig.~\ref{fig13}C;
errors based on bi-square are plotted in Fig.~\ref{fig13}D; see
Methods for details on linearity optimization).

Next, we tested a multiple exponential fitting of the data. The
rationale is that two exponential processes may represent
differences in two populations of cells, for example excitatory and
inhibitory cells. The fit resulting from a ``sum of exponential
processes'' was extremely good in minimum residual and reliable
prediction bounds for the data (Fig.~\ref{fig13}E). This ``sum of
exponential'' model (``exp2'') gave a very good performance in both
full length (dark blue) and reduced above ``$X_{min}$'' (red). The
``simple exponential'' model (exp1) reaches a very good fit only
for the reduced set (cyan) but not for the full length of the
avalanches (light brown). For comparison of ``exp1'' and ``exp2''
on different spike avalanches, with and without ``linearity
improvement'', see Fig.~\ref{fig13}F. Overall, it seems that both
exp1 and exp2 exhibit comparably high values of goodness of fit for
the reduced sets. However, only the double exponential fit was able
to fit the entire dataset.


\section*{Discussion}

In the present paper, we have analyzed and compared the avalanche
dynamics obtained from multielectrode recordings of spikes and
LFPs, for three species, cat, monkey and human. In each case, we
used recordings exclusively made in non-anesthetized brain states,
including quiet and active wakefulness, SWS (slow-wave sleep) and
REM (Rapid eye movement). The primary result of our analysis is
that there is no power-law scaling of neuronal firing, in any of
the examined recordings, including ``desynchronized'' EEG
states (wakefulness), SWS, and REM sleep. All species consistently
showed distributions which approached exponential distributions. 
This confirms previous findings of the absence of power-law
distributions from spikes in cats (Bedard et al., 2006), and
extends these findings to monkeys and humans. An obvious
criticism to that prior study is that a set of 8 electrodes is too low to
properly cover the system, and the absence of power-law may be due
to this subsampling. We show here that the same results are
obtained when a significantly higher density of recording is used,
confirming the absence of power-law.

In contrast, avalanche dynamics built from nLFPs displayed more
nuanced results. In some cases, the avalanche size distributions
appear to draw a straight line in log-log representations, but the
more reliable CDF-based tests did not show clear evidence for
power-law scaling. Indeed, statistical tests such as the KS test
did not give convincing evidence that these data are universally
distributed according to a power-law. More importantly, while nLFP
are related to firing activity, we showed that a similar behavior
was also observed for pLFP peaks. The avalanche analysis from
positive peaks displayed similar results as for negative peaks,
although positive peaks displayed a weaker statistical relation to
firing activity. Using 4 types of control/randomization we provide
very robust evidence that the fundamental differences between nLFP and pLFP are
not attributable to random behavior of spikes or LFP peaks. Yet still, the
discretized thresholded LFPs, show strikingly similar behavior in their
avalanche statistics. These findings render any conclusions about self-organized
criticality based on simple power-laws of PDFs as phenomenological. Together,
these results suggest that the
power-law behavior observed previously in awake monkey (Petermann
et al., 2009, Ribeiro 2010) cannot be reproduced in awake humans'
temporal cortex or cat and monkey motor cortex. This
conclusion also extends to slow-wave sleep and REM sleep, which we
found did not display power-law distributed avalanches, as defined
from either spikes or LFPs. In searching for the linear
domains in CDF based on the KS test, one can force the scaling exponent to fall
within the range of the plausible values
(comparable to those observed in known physical phenomena). Doing
so, of course, yields more conservative values of scaling, but means that such
scaling would be applicable to only a
limited range of data. In fact, unless the system has universal
scaling, there is always a tradeoff between the range to which a
scaling exponent can be extended (i.e. the linear regime in the
data) and the proximity of the scaling exponent value to those of a
narrow range (in this case, values of the SOC systems are of
interest). Our tests, did not force the scaling exponent to be
limited to values between 1-2, therefore it had a more stringent
emphasis on the linearity of more decades of the avalanche sizes.
In some cases where the data showed statistically significant
linearity, the obtained scaling exponents were an order of
magnitude higher than what falls in the range of the critical
regime of known physical phenomena. Conversely, these observations imply
that, a single scaling exponent is not sufficient to
explain the complex dynamics of ensemble activity. 

A possibility worth exploring is that some form of power-law in
LFPs is the result of volume conduction associated with LFPs
recorded in high density arrays. When a peak is detected, it is
often also present in many different channels. A possibility worth
to explore is whether the same event could be volume-conducted
across many channels in the array, which may lead to an artificial
increase the large-size avalanches. This possibility should be
examined by mathematical models of the volume conduction effect.  

It must be noted that the evidence for self-organized
criticality in neuronal cultures or in slices (Beggs and Plenz,
2003), as well as in anesthetized states (Hahn et al., 2010) is not
contradictory with the present findings. The wiring of {\it in
vitro} preparations, as well as the network dynamics in anesthesia,
are evidently different than in the intact brain. We find here that
there is no evidence for SOC in wakefulness and natural sleep
states, and for 3 different species.  On the other hand, the report
of power-law scaling of nLFPs avalanches in awake monkey (Petermann
et al., 2009) seems in contradiction with the present findings. 
Many possibilities exist to reconcile these observations, such as
differences between brain region, recording method, cortical layer
or volume conduction effects. These possibilities should be
investigated in future studies. Moreover, in a recent report
(Friedman 2012), it has been shown that data from high density
recordings (up to 512 electrodes) from from neural culture show
elements of universality and that avalanches can be collapsed into
a universal scaling function (Papanikolaou 2011). Such findings
confirm that brain circuits {\it in vitro} operate near criticality.
Further studies should examine how to reconcile such evidence with
the present {\it in vivo} findings.

Due to the high dimensionality of neural data, it is crucial
to separate the features of the inferred models that are induced
solely by the inference scheme from those that reflect natural
tendencies of the studied system (Mastromatteo and Marsili, 2011). 
In some cases, one could fit the data with different lines by
limiting the range of the decades within which a fit is analyzed. 
While it is indeed possible, and highly likely, that neural data at
this level follow a multi-scale regime, albeit such a property would
push the system away from cohesively operating at self organized
criticality because the relation between microscopic interaction of
the (neural) elements and collective behavior (of the cortical
network) no longer manifests in single valued features, like a
single scaling exponent.

Finally, it is important to emphasize that the present results were
obtained using statistical tests similar to previous statistical
analyses (Newman, 2005; Clauset et al., 2009). In particular, the
use of the CDF distribution rather than simple log-log
representations of the size distribution is a particularly severe
test to identify if a system scales as a power-law. The use of
statistical measures such as the Kolmogorov-Smirnov test
(Table.~\ref{tab1}, ~\ref{tab2} and ~\ref{tab3}) also constitutes a
good quantification of which distribution fits the data, and is
largely superior to the least square fit in double logarithmic
scale (Clauset et al., 2009). The uncertainty and goodness of fit
were estimated by 1000 repetitions of each fitted distribution. 
{We also showed that setting bounds to the fit can introduce
biases in favor of power-law fits, as analyzed previously (Clauset
et al., 2009).  In agreement with this, it was found with bounded
fits that power-law provides a better match to data compared to
exponential distributions (Klaus et al., 2011).  Our analysis shows
that neither power-law nor exponential distributions provide
acceptable fits to the datasets analyzed here.  Multi-exponential
fits suggest that bi-exponential processes provide a particularly
good fit to the distributions, which suggests that the underlying
neuronal dynamics is most compatible with two exponential
processes, which could be for example excitation and inhibition,
both scaling as exponential distributions.  Such a possibility
should be tested by further studies, and seem in agreement with the
complementary excitatory and inhibitory dynamics found in the awake
and sleeping brain (Peyrache et al., 2012).}



\section*{Acknowledgments}

Research supported by Centre National de la Recherche Scientifique
(CNRS, France), Agence Nationale de la Recherche (ANR, France),
European Community Future and Emerging Technologies program
(BRAINSCALES grant FP7-269921), the National Institutes of Health
(NIH grants 5R01NS062092, R01 EB009282) and DARPA (BAA05-26,
Revolutionizing Prosthetics). N.D. is supported by a fellowship
from Ecole de Neurosciences de Paris (ENP).


\section*{References}
\bibliography{template}
{Bak 1987}, Bak, P., Tang, C. , Wiesenfeld, K. Self-organized criticality: An
explanation of
the 1=f noise. Phys. Rev. Lett. 59, 381-384 (1987).
\\{Bak 1995}, Bak P, Paczuski M. Complexity, contingency, and criticality. PNAS
1995.
\\{Bak 1996},P. Bak, How Nature Works (Springer-Verlag, New York,1996).
\\{Bauke 2007},H. Bauke, Parameter estimation for power-law tail distributions
by maximum likelihood methods,
Eur. Phys. J. B, 58 (2007), pp. 167–173.
\\{Bedard 2006}, Bedard, C., Kroger, H., , Destexhe, A. (2006). Does the 1/ f
frequency-scaling of brain signals reflect self-organized critical states?
Physical Review Letters, 97, 118102.
\\{Bedard 2006b}, Bedard, C., Kroger, H., , Destexhe, A. (2006). Model of
low-pass filtering of local field potential in brain tissue. Phys Rev E Stat
Nonlin Soft Matter Phys, 73:051911.
\\{Begs 2003},Beggs JM, Plenz D (2003) Neuronal avalanches in neocortical
circuits. J Neurosci 23:11167–11177
\\{Chialvo 2010}, Chialvo D, Emergent complex neural dynamics, Nature Physics,
2010.
\\{Clauset 2009}, A. Clauset, C.R. Shalizi, and M.E.J. Newman, "Power-law
distributions in empirical data" SIAM Review 51(4), 661-703 (2009)
\\{Frette 1996}. Avalanche dynamics in a pile of rice. Nature. 1996
\\ {Friedman 2012} Friedman N, Ito S, Brinkman BA, Shimono M, Lee DeVille RE,
Dahmen KA, Beggs JM, Butler TC. Universal critical dynamics in high resolution
neuronal avalanche data. Physical Review Letters, (in press, 2012).
\\{Giesinger2001} Giesinger, T. Scale invariance in biology: 
coincidence or footprint of a universal mechanism? Biol. Rev.
 76, 161-209 (2001).
\\{Goldstein 2004}, M. L. Goldstein, S. A. Morris, and G. G. Yen, Problems with
fitting to the power-law
distribution, Eur. Phys. J. B, 41 (2004), pp. 255–258.
\\{Hahn 2010}, Hahn G, Petermann T, Havenith MN, Yu S, Singer W, et al. (2010)
Neuronal
avalanches in spontaneous activity in vivo. J Neurophysiol.
\\{Hahn 2011}, Hahn G, Monier C and Fr\'egnac Y.  Revisiting power law in vivo
as
a function of the global brain state, using multiple recordings in
anesthetized cat V1.  {\it Soc. Neurosci. Abstracts} 451.10, 2011.
\\{Jensen 1998},H. J. Jensen, Self-Organized Criticality: Emergent Complex
Behavior in Physical and Biological Systems (Cambridge University Press,
Cambridge, England, 1998).
\\{Klaus 2011}. Klaus A, Yu S, Plenz D (2011) Statistical Analyses Support Power
Law Distributions Found in Neuronal Avalanches. PLoS ONE 6(5): e19779
\\{Malamud 1988},Malamud, B. D., Morein, G. , Turcotte, D. L. Forest fires: An
example of
self-organized critical behaviour. Science 281, 1840-1842 (1998).
\\{Mastromatteo 2011} Mastromatteo I, Marsili M. On the
Criticality of Inferred Models. Journal of Statistical Mechanics: Theory and
Experiment 2011, no. 10 (October 10, 2011): P10012.
\\{Mersenne 1998} Matsumoto, M.; Nishimura, T. (1998). "Mersenne twister: a
623-dimensionally equidistributed uniform pseudo-random number generator". ACM
Transactions on Modeling and Computer Simulation 8 (1): 3–30.
\\{Newman 2005}. Newman, Power laws, Pareto distributions and Zipf's law
Contemporary Physics 46, 323-351 (2005)
\\{Papanikolaou 2011} Papanikolaou S, Bohn P, Sommer RL, Durin R, Zapperi Z, and
Sethna JA. “Universality Beyond Power Laws and the Average Avalanche Shape.”
Nature Phys. 7, no. 4 (April 2011): 316–320.
\\ {Parker 2011}. Parker RA, Davis TS, House PA, Normann RA, Greger B. The
functional consequences of chronic, physiologically effective intracortical
microstimulation. in: J. Schouenborg, M. Garwicz and N. Danielsen (Eds.)
Progress in Brain Research, Vol. 194. Brain Machine Interface (2011).
\\{Petermann 2009} Petermann, T. Thiagarajan TC, Lebedev MA, Nicolelis MA,
Chialvo DR, et al. Spontaneous cortical activity in awake monkeys composed
of neuronal avalanches. Proc. Natl Acad. Sci. USA 106, 15921-15926 (2009).
\\{Peters 2006},Peters, O. , Neelin, D. Critical phenomena in atmospheric
precipitation.
Nature Phys. 2, 393-396 (2006).
\\{Peyrache 2012}, Adrien Peyrache, Nima Dehghani, Emad Eskandar, Joseph Madsen,
William Anderson, Jacob Donoghue, Leigh R Hochberg, Eric Halgren, Sydney S. Cash
and Alain Destexhe, Spatio-temporal dynamics of neocortical excitation and
inhibition during human sleep., PNAS : , (2012) 
\\{Press 1992a} Press, W. H.; Flannery, B. P.; Teukolsky, S. A.; and Vetterling,
W.T. "Random Numbers." Ch. 7 in Numerical Recipes: The Art of Scientific
Computing, 3rd ed. Cambridge, England: Cambridge University Press, pp.
266-306, 1992.
\\{Press 1992b} Press, W. H.; Flannery, B. P.; Teukolsky, S. A.; and Vetterling,
W.T. "Nonlinear models." Ch. 15 in Numerical Recipes: The Art of Scientific
Computing, 3rd ed. Cambridge, England: Cambridge University Press, pp.
266-306, 1992.  
\\{Ribiero 2010}, Ribeiro T, Copelli M, Caixeta F, Belchior H, Chialvo DR,
Nicolelis M, Ribeiro S. Spike Avalanches Exhibit Universal Dynamics across the
Sleep-Wake Cycle. Plos One 2010.
\\{Steriade 2001}. Steriade M. The intact and sliced brain. 
\\{Toboul 2010} Touboul, J. , Destexhe, A. Can power-law scaling and neuronal
avalanches
arise from stochastic dynamics? PLoS One 5, (2010).
\\{Truccolo 2010} Truccolo, W., Hochberg, L., Donoghue, J. (2010). Collective
dynamics in human and monkey sensorimotor cortex: Predicting single neuron
spikes. Nature Neuroscience, 13, 105–111.


\clearpage


\clearpage

\section*{Tables}
\begin{table}[!ht]
\caption{\bf{Summary spike avalanche}}
\begin{center}
\begin{tabular}{llllll}
\textbf{\textit{Species}} & \textbf{\textit{Loc}} & \textbf{\textit{State}} &
\textbf{\textit{CDF exponent}} & \textbf{\textit{Pval}} &
\textbf{\textit{gof}}\\
Monkey i & MI & Awake & 3.4413  $\pm$   0.7616 & 0.0419 $\pm$   0.1152 & 0.0442
$\pm$ 0.0216\\
Monkey i & Pmd & Awake & 4.1660  $\pm$  0.6590 & 0.1130  $\pm$  0.2140 & 0.0180 
$\pm$ 0.0050\\
Monkey ii & MI & Awake & 4.6250 $\pm$   0.4730 & 0.4550  $\pm$  0.3600 & 0.0330 
$\pm$ 0.0120\\
Monkey iii & MI & SWS & 4.5560 $\pm$   0.7980 & 0.0030 $\pm$  0.0100 & 0.0220
$\pm$  0.0080\\
Monkey iii & Pmd & SWS & 3.7760  $\pm$  0.8660 & 0    $\pm$     0 & 0.0430 
$\pm$ 
0.0170\\
Cat ii & MI & Awake & 2.8412   $\pm$ 1.2184 & 0.3056 $\pm$   0.3844 &
0.0599 $\pm$  0.0368\\
Cat iii & Parietal & Awake & 3.1410   $\pm$ 0.8720 & 0.2010 $\pm$   0.3680 &
0.0270 $\pm$  0.0180\\
Cat iii & Parietal & SWS & 4.2110 $\pm$   0.7930 & 0.3290  $\pm$  0.3620 &
0.0350
$\pm$  0.0140\\
Cat iii & Parietal & REM 1 & 3.3240  $\pm$  0.8150 & 0.2990  $\pm$  0.2170 &
0.0290 
$\pm$ 0.0110\\
Cat iii & Parietal & REM 2 & 3.4050  $\pm$  0.8250 & 0.4250  $\pm$  0.4470 &
0.0230 
$\pm$ 0.0140\\
Human i & Temporal & Awake & 3.5490   $\pm$ 0.8790 & 0.3870  $\pm$  0.3650 &
0.0210 $\pm$  0.0080\\
Human i & Temporal & SWS 1 & 3.6340  $\pm$  0.6410 & 0.3790 $\pm$   0.3150 &
0.0250 $\pm$  0.0100\\
Human i & Temporal & SWS 2 & 3.2550 $\pm$   0.5770 & 0.1710  $\pm$  0.2670 &
0.0330 $\pm$  0.0150\\
Human i & Temporal & REM 1 & 3.3740  $\pm$  0.8560 & 0.0930 $\pm$   0.1720 &
0.0300 $\pm$  0.0090\\
Human i & Temporal & REM 2 & 3.6430 $\pm$   0.5540 & 0.0960 $\pm$   0.1950 &
0.0320 $\pm$  0.0170\\
Human i & Temporal & Awake & 3.9200  $\pm$  0.7970 & 0.0080 $\pm$   0.0230 &
0.0090 $\pm$  0.0070\\
Human i & Temporal & SWS & 3.8950  $\pm$  0.7630 & 0.0070 $\pm$   0.0140 &
0.0100 $\pm$  0.0070
\end{tabular}
\end{center}
\begin{flushleft}Cross species summary of spike avalanche\end{flushleft}
\label{tab1}
\end{table}
\begin{table}[!ht]
\caption{\bf{Detailed Awake spike Avalanche}}
\begin{center}
\begin{tabular}{ccccc}
\textbf{\textit{Loc}} & \textbf{\textit{Bin size(ms)}} & \textbf{\textit{CDF
exponent}} & \textbf{\textit{Pval}} & \textbf{\textit{gof}}\\
MI & 1 & 2.5 & 0 & 0.036\\
MI & 3 & 5 & 0.008 & 0.020\\
MI & 5 & 3.36 & 0 & 0.029\\
MI & 7 & 3.63 & 0 & 0.039\\
MI & 9 & 3.03 & 0 & 0.047\\
MI & 11 & 3.83 & 0.327 & 0.034\\
MI & 13 & 3.35 & 0 & 0.060\\
MI & 15 & 2.83 & 0 & 0.089\\
PMd & 1 & 4.1 & 0 & 0.006\\
PMd & 3 & 2.81 & 0 & 0.021\\
PMd & 5 & 5 & 0 & 0.018\\
PMd & 7 & 4.85 & 0.061 & 0.017\\
PMd & 9 & 4.03 & 0 & 0.022\\
PMd & 11 & 4.21 & 0.018 & 0.024\\
PMd & 13 & 4.25 & 0.216 & 0.019\\
PMd & 15 & 4.08 & 0.61 & 0.017
\end{tabular}
\end{center}
\begin{flushleft}Monkey i detailed table.\end{flushleft}
\label{tab2}
\end{table}
\begin{table}[!ht]
\caption{\bf{Detailed Awake LFP Avalanche}}
\begin{center}
\begin{tabular}{cccccc}
\textbf{\textit{Bin size(ms)}} & \textbf{\textit{Polarity}} &
\textbf{\textit{Threshold}} & \textbf{\textit{CDF exponent}} &
\textbf{\textit{Pval}} & \textbf{\textit{gof}}\\
1 & neg & Low & 1.71 & 0 & 0.019\\
3 & neg & Low & 2.99 & 0.056 & 0.051\\
5 & neg & Low & 2.55 & 0 & 0.052\\
7 & neg & Low & 2.84 & 0.074 & 0.052\\
9 & neg & Low & 2.42 & 0 & 0.053\\
11 & neg & Low & 2.37 & 0 & 0.059\\
13 & neg & Low & 2.43 & 0 & 0.054\\
15 & neg & Low & 2.36 & 0 & 0.052\\
1 & neg & Mid & 1.83 & 0.002 & 0.015\\
3 & neg & Mid & 2.79 & 0.425 & 0.040\\
5 & neg & Mid & 2.84 & 0.55 & 0.042\\
7 & neg & Mid & 2.81 & 0.376 & 0.048\\
9 & neg & Mid & 2.84 & 0.345 & 0.050\\
11 & neg & Mid & 2.84 & 0.435 & 0.048\\
13 & neg & Mid & 2.71 & 0.098 & 0.058\\
15 & neg & Mid & 2.74 & 0.204 & 0.056\\
1 & neg & High & 1.9 & 0 & 0.018\\
3 & neg & High & 1.55 & 0 & 0.029\\
5 & neg & High & 2.44 & 0.645 & 0.036\\
7 & neg & High & 2.43 & 0.201 & 0.046\\
9 & neg & High & 2.41 & 0.672 & 0.036\\
11 & neg & High & 2.39 & 0.67 & 0.035\\
13 & neg & High & 2.3 & 0.496 & 0.036\\
15 & neg & High & 2.3 & 0.36 & 0.040\\
1 & pos & Low & 1.68 & 0 & 0.020\\
3 & pos & Low & 1.37 & 0 & 0.073\\
5 & pos & Low & 3.03 & 0 & 0.066\\
7 & pos & Low & 4.21 & 0.762 & 0.051\\
9 & pos & Low & 3.59 & 0.585 & 0.048\\
11 & pos & Low & 3.39 & 0.43 & 0.047\\
13 & pos & Low & 2.98 & 0.079 & 0.046\\
15 & pos & Low & 2.9 & 0.032 & 0.052\\
1 & pos & Mid & 1.74 & 0 & 0.018\\
3 & pos & Mid & 3.67 & 0.128 & 0.062\\
5 & pos & Mid & 3.79 & 0.047 & 0.069\\
7 & pos & Mid & 5 & 0.827 & 0.061\\
9 & pos & Mid & 3.78 & 0.797 & 0.041\\
11 & pos & Mid & 3.68 & 0.926 & 0.036\\
13 & pos & Mid & 3.87 & 0.797 & 0.049\\
15 & pos & Mid & 3.51 & 0.553 & 0.046\\
1 & pos & High & 1.76 & 0.009 & 0.020\\
3 & pos & High & 1.47 & 0 & 0.061\\
5 & pos & High & 3.19 & 0.169 & 0.067\\
7 & pos & High & 3.17 & 0.063 & 0.066\\
9 & pos & High & 3.07 & 0.251 & 0.061\\
11 & pos & High & 3.09 & 0.325 & 0.059\\
13 & pos & High & 3.18 & 0.286 & 0.062\\
15 & pos & High & 2.74 & 0.033 & 0.061
\end{tabular}
\end{center}
\begin{flushleft}Human i detailed Table.\end{flushleft}
\label{tab3}
\end{table}

\end{document}